\documentclass[prb,reprint,11pt]{revtex4-1}

\usepackage[T1]{fontenc} % Use modern font encodings
\usepackage{graphicx}
\usepackage{textcomp}
\usepackage[english]{babel}
\usepackage[utf8]{inputenc}
\usepackage[T1]{fontenc}
\usepackage{hyperref}
\usepackage{amsmath}
\usepackage{amssymb}
\usepackage[amssymb]{SIunits}
\usepackage{physics}
\usepackage{xcolor}
\usepackage{mciteplus}

\newcommand{\UPDAddress}{Laboratoire Matériaux et Phénomènes Quantiques, Université Paris Diderot, 
Sorbonne Paris Cité, CNRS-UMR 7162, 75013 Paris, France}
\newcommand{\LeedsAddress}{School of Electronic and Electrical Engineering, University of Leeds, LS2 9JT Leeds, United Kingdom}

\begin{document}

\title{Ultra-Strong Light-Matter Coupling in Deeply Subwavelength THz LC resonators}

\author{Mathieu Jeannin}
\email{mathieu.jeannin@univ-paris-diderot.fr}
%\altaffiliation{A shared footnote}
\author{Giacomo Mariotti Nesurini}
\author{Stéphan Suffit}
\author{Djamal Gacemi}
\author{Angela Vasanelli}
\affiliation{\UPDAddress}

\author{Lianhe Li}
\author{Alexander Giles Davies}
\author{Edmund Linfield}
\affiliation{\LeedsAddress}
\author{Carlo Sirtori}
\author{Yanko Todorov}
\affiliation{\UPDAddress}

\begin{abstract}
The ultra-strong light-matter coupling regime has been demonstrated in a novel three-dimensional 
inductor-capacitor (LC) circuit resonator, embedding a semiconductor 
two-dimensional electron gas in the capacitive part. The fundamental resonance of the LC circuit interacts with the 
intersubband plasmon excitation of the electron gas at $\omega_c = 3.3$~THz with a normalized coupling strength 
$2\Omega_R/\omega_c = 0.27$.  Light matter interaction is driven by the quasi-static electric field 
in the capacitors, and takes place in a highly subwavelength effective volume 
$V_{\mathrm{eff}} = 10^{-6}\lambda_0^3$ . This enables the observation of the ultra-strong light-matter 
coupling with $2.4\times10^3$ electrons only. Notably, our fabrication protocol can be applied to the integration 
of a semiconductor region into arbitrary nano-engineered three dimensional meta-atoms. 
This circuit architecture can be considered 
the building block of metamaterials for ultra-low dark current detectors. 
\end{abstract}

\maketitle

%%\section{Introduction}
Metamaterials were introduced to enable new electromagnetic properties of matter which are not naturally 
found in nature.
Celebrated examples of such achievements are, for instance, negative refraction \cite{shelby_experimental_2001} 
and artificial magnetism.\cite{pendry_magnetism_1999} 
The unit cells of metamaterials are artificially designed meta-atoms that have dimensions ideally much smaller 
than the wavelength of interest $\lambda_0$.\cite{cai_optical_2010} 
Such meta-atoms act as high frequency inductor-capacitor (LC) resonators which sustain a resonance 
close to $\lambda_0\propto\sqrt{LC}$.\cite{cai_optical_2010} 
The resonant behaviour, occurring into highly subwavelength volumes, generates high electromagnetic 
field intensities which, as pointed out by the seminal paper of \citeauthor{pendry_magnetism_1999},
\cite{pendry_magnetism_1999} are crucial to implement artificial electromagnetic properties of a 
macroscopic ensemble of meta-atoms. 
Moreover, the ability to control and enhance the electromagnetic field at the nanoscale is beneficial 
for optoelectronic devices, such as nano-lasers \cite{wang_unusual_2017} electromagnetic sensors
\cite{chen_metamaterials_2012,belacel_optomechanical_2017,alves_mems_2018} and detectors.
\cite{shrekenhamer_experimental_2012,viti_nanowire_2014,luxmoore_graphenemetamaterial_2016,palaferri_ultra-subwavelength_2016,palaferri_room-temperature_2018} 
For instance, metamaterial architectures have lead to a substantial decrease of the 
thermally excited dark current in quantum infrared detectors, resulting in higher temperature operation.
\cite{palaferri_ultra-subwavelength_2016,palaferri_room-temperature_2018} 

The LC circuit can be seen as a quantum harmonic oscillator sustaining vacuum electric field fluctuations that scale as 
$1/V_{\mathrm{eff}}^{1/2}$, where $V_{\mathrm{eff}}$ is the effective volume of the capacitive parts.
\cite{louisell_quantum_1990}  
For an emitter/absorber inserted between the capacitor plates, the light-matter interaction is proportional to 
$1/V_{\mathrm{eff}}^{1/2}$, and thus strongly enhanced.
Fundamental electro-dynamical phenomena, such as the Purcell effect\cite{purcell_resonance_1946} 
and strong light-matter coupling regime\cite{yamamoto_semiconductor_2000} can therefore be observed. 
In the strong coupling regime, energy is reversibly exchanged between the matter excitation 
and the electromagnetic resonator at the Rabi frequency $\Omega_R$. 
This results in an energy splitting of the circuit resonance into two polaritons states separated by $2\hbar\Omega_R$. 
The regime of strong coupling has been observed in many physical systems which have been reviewed 
e.g. in Refs. \citenum{baranov_novel_2018,frisk_kockum_ultrastrong_2019,forn-diaz_ultrastrong_2018}, 
and some specific realizations with metamaterial resonators were achieved in the sub-THz,
\cite{scalari_ultrastrong_2012} 
THz\cite{todorov_ultrastrong_2010,strupiechonski_sub-diffraction-limit_2012,dietze_ultrastrong_2013} 
and the Mid-IR\cite{benz_control_2015,askenazi_ultra-strong_2014,askenazi_midinfrared_2017} part of the spectrum. 
In these systems, the highly subwavelength interaction volumes combined with the collective effect of $N_e$ identical 
electronic transitions result into high coupling constants $\Omega_R\propto(N_e/V_{\mathrm{eff}})^{1/2}$, 
and allow reaching the ultra-strong coupling regime where the Rabi splitting becomes of the same order of magnitude 
as the frequency of the material excitation $\tilde{\omega}$, $2\Omega_R/\tilde{\omega}\approx1$.
\cite{ciuti_quantum_2005} 
Since there is virtually no lower limit for the interaction volume $V_{\mathrm{eff}}$ in LC resonators, 
the fascinating regime of ultra-strong coupling can be realized in structures having few electrons only.
\cite{todorov_few-electron_2014,keller_few-electron_2017} 
In such limit, the effective bosonization procedure employed to describe the properties of the two-dimensional 
electron gas breaks down, and one can investigate the unique regime where the few electrons in the system have 
to be exactly treated as fermions.\cite{todorov_few-electron_2014} 

Recently, the ultra-strong coupling regime with a small number of electrons has been experimentally observed 
by coupling transitions between Landau levels in a two dimensional electron gas under a high magnetic field 
and nanogap complementary bow-tie antennas, with a record low number of 80 electrons.\cite{keller_few-electron_2017}  
Those studies were performed in the sub-THz part of the spectrum (300GHz) using resonators based on a planar geometry. 
Here, we demonstrate a three-dimensional metamaterial architecture that has the potential to go beyond this limit 
in the THz range (3THz), without the need for a magnetic field. 
Our metamaterial allows confining the electric field in all directions of space into nanoscale volumes, 
on the order of $V_{\mathrm{eff}}=10^{-6}\lambda_0^3$. The resonance of the structure is coupled 
to an intersubband (ISB) transition of high density electron gas in the ground state of semiconductor quantum wells 
(QWs). A relative Rabi frequency of $2\Omega_R/\tilde{\omega}=0.27$ is attained with a record low 
overall number of electrons $N_e\approx10^3$ for intersubband systems. 
Other reports on coupled LC resonators-ISB transitions systems in the THz spectral range reached 
similar coupling constants of $2\Omega_R/\tilde{\omega}=0.2$ with a much greater mode volume and electron number 
($V_{\mathrm{eff}}\approx10^{-5}\lambda_0^3$ and $N_e=4.6\times 10^4$).\cite{paulillo_room_2016} 
Comparable number of coupled electrons have been reported in the 
MIR spectral range\cite{benz_strong_2013, malerba_towards_2016} using very small mode volume cavities 
with $V_{\mathrm{eff}}\approx 6-9 \times 10^{-6}\lambda_0^3$, but at the expense of reducing the coupling constant 
($2\Omega_R/\tilde{\omega}=0.1$ in Ref.~\citenum{benz_strong_2013} and 
$2\Omega_R/\tilde{\omega}=0.05$ in Ref.~\citenum{malerba_towards_2016}). 
We use the dependence of the polariton splitting on the effective mode volume as a near field probe to 
estimate the highly subwavelength volume of our resonators in comparison with reference microcavity systems.
\cite{todorov_ultrastrong_2010,feuillet-palma_extremely_2012} 
These results are obtained thanks to a novel fabrication process that allows structuring metal patterns on both sides 
of a very thin semiconductor layer, which opens many degrees of freedom in the design and 
functionalization of 3D metamaterial architectures into optoelectronic devices.  

%\section{System presentation and samples}

\begin{figure*}[!ht]
\centering
\includegraphics[width=\linewidth]{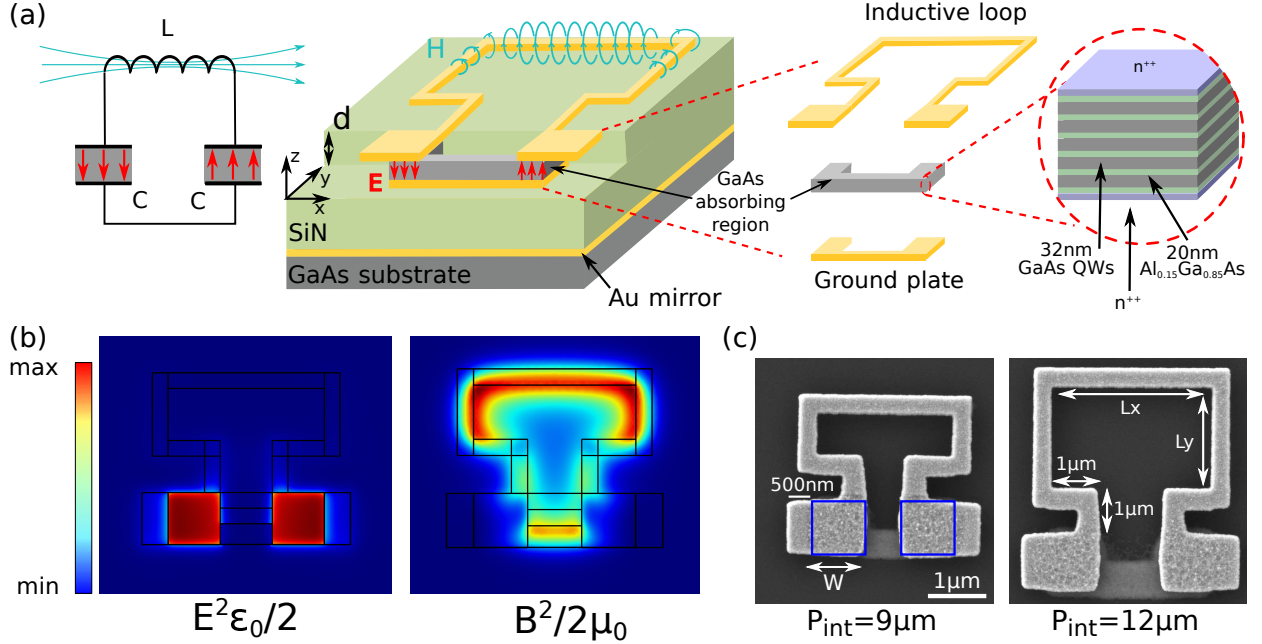}
\caption{
Sketch of the THz LC resonator and resonator mode. 
(a) The absorbing region (GaAs/AlGaAs QWs, see the Supplementary Information for a detailed composition and 
band structure) is embedded inside the resonator capacitors, defined by the overlap of two metallic plates. 
The device effectively acts as a LC resonator, where the magnetic field loops around the top wire, 
and the electric field is confined between the metallic plate and normal to their surface. 
This is further evidenced by numerical simulations (b) showing the electric (left) and magnetic (right) energy density. 
We can see that the electric field is well confined in the capacitors, while the magnetic energy follows the 
inductive loop. The two fields are spatially disentangled. 
The host substrate is coated with an Au mirror for reflectivity experiments, 
and the SiN spacer thickness is kept large enough (3~$\mu$m) to prevent any coupling between the resonator and 
the mirror. 
(c) Scanning electron microscope image of two LC resonators with different internal perimeters.  
The blue squares on the left panel mark the capacitor area (1$\mu$m$\times$1$\mu$m), showing the 500~nm extensions 
ensuring an improved coupling with free-space radiation (see text).
}
\label{fig:LCPresentation}
\end{figure*}

Our THz LC resonator is introduced in Fig.~\ref{fig:LCPresentation}~(a), along with a sketch of the equivalent 
circuit. The bottom metallic ground plate is formed by two square capacitor plates of width $W$, connected by a thin 
stripe. The top metallic part is composed of two rectangular capacitor plates ($W \times W +0.5\mu$m) connected by a 
bent wire. Two parallel plate capacitors are thus formed at the overlap between the metallic pads, while the bent wire acts 
as an inductive loop. 
Note that the top metal plates are 500~nm wider than the bottom one, resulting in a 500~nm wide 
extension over the outer parts of the capacitor, as shown in Fig.~\ref{fig:LCPresentation}~(c). 
The capacitors area is shown by the blue squares on the left image of this panel. 
The small extensions allow engineering the fringing fields between the two capacitor plates. 
By breaking the symmetry of the in-plane component of the electric field, they allow efficient coupling between the resonator 
mode and far-field radiation, as determined by previous work on SiN-based resonators.
\cite{mottaghizadeh_nanoscale_2017}  
The circuit-like mode of the structure which oscillates at the lowest frequency confines the electric field 
in the capacitive areas while the magnetic field loops around the inductive wire, 
\cite{todorov_three-dimensional_2015,mottaghizadeh_nanoscale_2017} 
as shown in Fig. \ref{fig:LCPresentation}~(b), calculated using finite-elements method simulations. They 
represent the electric (left) and the magnetic (right) energy density in the center $xy$ plane of the resonator. 
The matter resonance is provided by GaAs/AlGaAs QWs inserted inside the capacitive elements. 
Note that the $z$-component of the electric field is the only active component for ISB absorption. 
The resonator mode described above thus naturally fulfills the ISB polarization selection rule, 
which requires the electric field to be oriented along the growth axis of the QWs. 

To fabricate the LC resonator, the ground plate is first patterned using e-beam lithography 
and used as a mask to etch the absorbing region with an inductively-coupled plasma. 
The structure is then encapsulated in a 3~$\mu$m thick SiN layer, and the surface of the sample is metalized 
with an optically thick Au layer. 
The latter serves as a mirror which blocks the transmission into the substrate, such as the reflected beam contains only the 
spectral features of the metamaterial array.
The sample (grown on a GaAs substrate) is then flipped and transferred to a host substrate using an epoxy, 
and the growth substrate is selectively etched away, revealing the bottom of the patterned absorbing region. 
The top inductive loop and rectangular capacitor plates are then defined using e-beam lithography. 
Figure \ref{fig:LCPresentation}~(c) shows scanning electron microscope images of typical LC resonators. 
In the following, we keep the lateral size of the capacitor $W=1~\mu$m.
The internal perimeter P$_{int}=2L_y+L_x+4\mu$m of the inductive loop is varied from 9 to 14$\mu$m to tune the 
resonant frequency across the 2-6~THz spectral region, as explained further. 
A single resonator fits in a square with a diagonal of 4.2~$\mu$m (P$_{int}$=9~$\mu$m) to 7~$\mu$m 
(P$_{int}$=12~$\mu$m), much smaller than the vacuum wavelength $\lambda_0$=100~$\mu$m. 
The footprint of a single resonator ranges from $3\times3$ to 5$\times$5~$\mu$m$^2$, e.g. $10^{-3}\lambda_0^2$ to 
$2.5\times 10^{-3}\lambda_0^2$. 

\begin{figure*}[!ht]
\centering
\includegraphics[width=\linewidth]{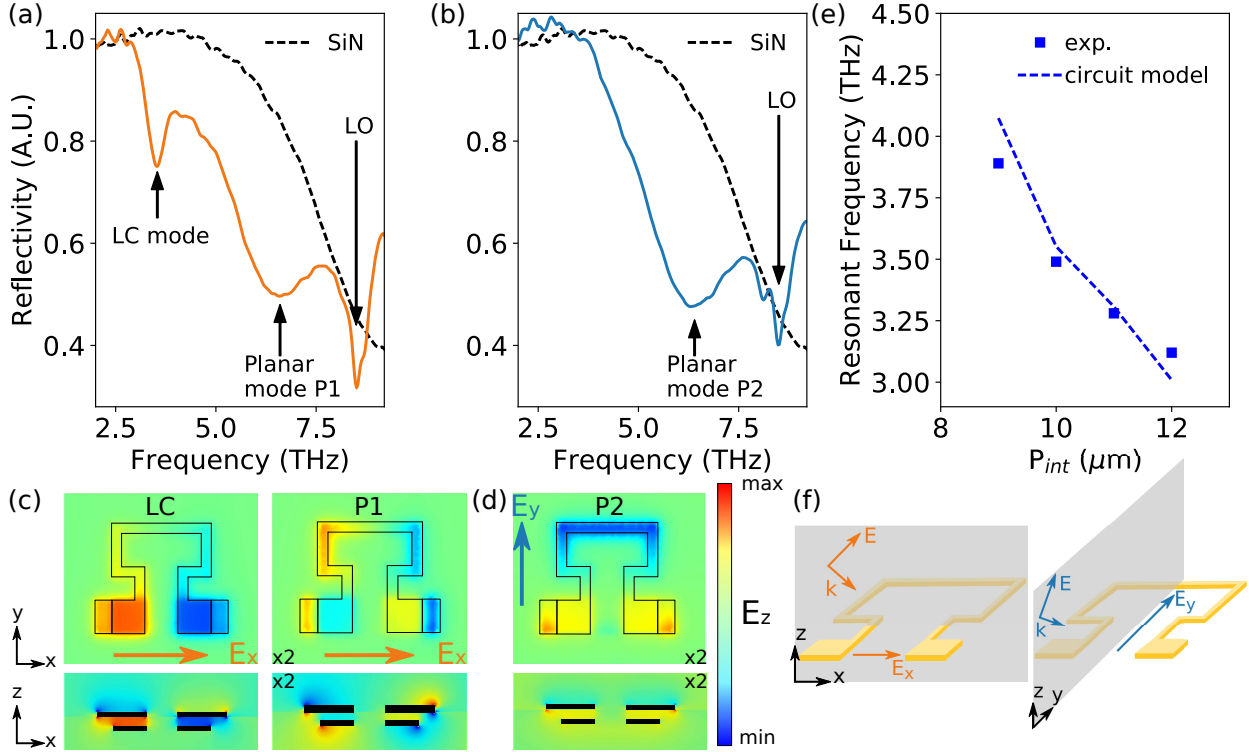}
\caption{Reflectivity of a LC resonator array at 45° of angle of incidence in TM configuration using light polarized 
(a) along the capacitor line (b) normal to the capacitor line (see (c) and (d)) and reflectivity of the SiN-Au stack 
(black dashed line). 
(c) $E_z$ component of the electric field obtained from numerical simulations for the two modes with an 
incident light polarized along the $x$ direction in a $x-y$ (top) and a $x-z$ (bottom) cut plane 
at the center of the capacitors. Note that for the planar mode, the magnitude of the $E_z$ has been multiplied by 2. 
(d) $E_z$ same as (c) for an electric field polarized along the $y$ direction. 
Note that the magnitude of the $E_z$ has been multiplied by 2.
(e) Blue squares: 
Resonant frequencies of a set of LC resonators with W=1~$\mu$m and $d$=290~nm, 
varying simultaneously $L_x$ and $L_y$ to increase 
P$_{int}$ while maintaining a square aspect ratio. 
Blue dashes: 
Resonant frequencies of the equivalent circuit. 
(f) Sketch of the optical configuration, ensuring a TM-polarized light excitation with respectively $E_x$ and $E_y$ 
in-plane electric field projection. The gray plane represents the plane of incidence ($x-z$ plane and $y-z$ plane respectively).
}
\label{fig:Reflectivity}
\end{figure*}

We first probe the optical properties of our system at room temperature, where the ISB absorption can be neglected 
as the thermal energy is sufficiently high to equally populate the first few energy levels of the QWs, 
and we can study the electromagnetic modes of the resonator alone.
\cite{todorov_polaritonic_2012} 
We perform reflectivity experiments using a dry-air purged Fourier Transform Spectrometer (FTIR) (Brucker Vertex 70v) 
and a globar source. The FTIR is equipped with a custom made reflectivity setup, which allows focusing the globar beam 
on the sample with the help of a pair of F1 parabolic mirrors, and the reflected beam is measured with a He cooled 
Ge bolometer (QMC Instruments).  Light is linearly polarized and impinging at 45° on the sample, as sketched in 
Fig.~\ref{fig:Reflectivity}~(f). 
To ensure good spatial overlap with the globar beam, we fabricate dense arrays composed of $\approx$50 000 resonators 
separated by 2~$\mu$m from each other, with a total area of 2$\times$2~mm$^2$. 
All spectra are normalized by the reference from a Au mirror. 
In Fig.~\ref{fig:Reflectivity}~(a)-(b) we show typical spectra obtained for resonators with P$_{int}=10$~$\mu$m, with light 
polarized respectively along the line formed by the capacitors ($E_x$), and orthogonally to that line ($E_y$). 
By comparing the two spectra, we first note the presence of a dip in the reflectivity spectrum for 
$E_x$-polarized light at 3.5~THz, which is absent in the $E_y$-polarized spectrum. 
This corresponds to the LC mode of the resonator represented in Fig.\ref{fig:LCPresentation}. 
A broad resonance is observed in both polarizations at 6.5~THz, whose origin will be discussed later. 
A strong dip is observed just above 8~THz in both polarizations. This corresponds to the lower frequency edge 
of the GaAs reststrahlen band arising from the optical phonons,\cite{kittel_introduction_1996} 
confirming the very strong localization of the electric field in the semi-conductor layer enabled by our device, 
as the GaAs only constitutes 3.5\% of the surface probed by the THz beam. 
The baseline in all spectra arises from the 3~$\mu$m thick SiN layer, as confirmed by the reflectivity spectra 
from an area without any patterns (dashed curves in Fig.~\ref{fig:Reflectivity}~(a)-(b)). 
Indeed the SiN layer has a residual absorption in this spectral region.\cite{cataldo_infrared_2012}

To clarify the origin of the various resonances observed in the experiments we simulated 
the reflectivity of the structure using a commercial finite element method software (Comsol v.5.3a). 
Maps of the $E_z$ component of the electric field for the three modes  
are shown in Figs.~\ref{fig:Reflectivity}~(c)-(d) in two cut planes located at the center of the capacitors. 
Note that in the right panel of Fig.~\ref{fig:Reflectivity}~(c) and in Fig.~\ref{fig:Reflectivity}~(d) 
the magnitude of the electric field $E_z$ has been multiplied by two for clarity. 
The $xy$-plane color maps show the expected electric field distribution for the LC mode, and 
reveal that the resonance at 6.5~THz actually corresponds to two different modes (P1 and P2) 
excited with different polarizations ($E_x$ and $E_y$ respectively). Comparing the $xz$-plane maps, 
we see that only the LC mode efficiently confines the electric field inside the capacitors. 
P1 and P2 are modes where the electric field lies mainly in the plane of the inductive loop, 
reminiscent of the modes observed in planar metamaterial resonators.\cite{zhou_magnetic_2007} 
Notably, simulations including the top loop alone 
(and not the ground plate) show the same P1 and P2 modes but not the three-dimensional LC resonance. 
These two modes couple to the isotropic phonon absorption in the GaAs regions below the Au pads, 
as they still have a spectral overlap with the GaAs reststrahlen band.
A detailed survey of the different modes sustained by the structure is beyond the scope of this paper, and from now on 
we will restrict the analysis to the LC mode, which is the only one providing electric field almost exclusively localized 
in the capacitive parts and satisfying the selection rule for the QW ISB absorption. 

The LC resonance is tuned by changing the internal perimeter P$_{int}$ of the inductive loop, 
while keeping the capacitance parts fixed. 
The resonant frequencies as a function of P$_{int}$ are reported in Fig.~\ref{fig:Reflectivity}~(e) (symbols). 
We compare the experimental results with those of a lumped element model which 
corresponds to the equivalent circuit sketched in 
Fig.~\ref{fig:LCPresentation}(a) (see Supplementary Materials for more information). 
The resonant frequencies of the model, provided by $\omega_{LC} = \sqrt{2/LC}$ are plotted 
in dashed lines in Fig.~\ref{fig:Reflectivity}~(c), and are in good agreement 
with those extracted from the measurements. 
The calculated inductance varies from $L=11$~pH to $L=6$~pH. The calculated capacitance is $C=480$~aF, 
which compares well with the value for an ideal parallel plate capacitor $C=\varepsilon \varepsilon_0 W^2/d=374$ aF, with 
$\varepsilon$ and $\varepsilon_0$ the material and vacuum permittivities. 
The difference between these two values can be explained by the contribution 
of the fringing fields and of the in-plane parasitic gap capacitance between the two metal pads. 
This evidences that the fundamental mode of our resonators operates in the near quasi-static limit. 

%\section{USC in first sample}

To explore the ultra-strong light-matter coupling regime, we have inserted  in our resonators
an absorbing region consisting of 5 repetitions of 32nm GaAs quantum wells separated by 20nm 
Al$_{0.15}$Ga$_{0.85}$As barriers, 
similar to the design described in Ref. \citenum{todorov_ultrastrong_2010,todorov_polaritonic_2012}. 
The QWs are modulation-doped by Si $\delta$-doped regions placed 5~nm away from the QW, 
with a nominal sheet carrier density of $4\times10^{11}$~cm$^{-2}$. In such thin double-metal structures 
the presence of the two metal-semiconductor interfaces creates a depletion layer 
on each side of the semiconductor region, bending the conduction band profile 
and usually depleting 2-3 QWs.\cite{sze_physics_2006, todorov_polaritonic_2012} 
To compensate for this effect, we introduced a doped AlGaAs spacing layer and a doped GaAs external layer 
(both doped at $2\times10^{18}$~cm$^{-3}$) on each side of the absorbing region, 
which is of total thickness $d=290$~nm. 
Additional details on the band structure and sample design are provided in the Supplementary materials. 
The matter excitation coupled to the LC resonator is an intersubband plasmon of frequency 
$\tilde{\omega}=\sqrt{\omega_{12}^2+\omega_{P}^2}$, 
where $\omega_{12}$ is the bare ISB transition frequency and $\omega_P$ is the plasma frequency of the 
two-dimensional electron gas in the QWs.\cite{Ando_electronic_1982} 
The latter is provided by the expression 
$\omega_P = \sqrt{(n_1-n_2)e^2/(\varepsilon \varepsilon_0m^{\ast}L_\mathrm{QW,eff})}$, 
where  $e$ is the electron charge, $m^{\ast}$ is the electron effective mass in the QW, $n_1$ (resp. $n_2$) 
is the surface electron density in the first (second) subband, and $L_\mathrm{QW,eff}$ is an effective length 
of the quantum well as defined in Ref. \citenum{todorov_intersubband_2012}.
%\bibnote{
%In Ref.~\citenum{todorov_intersubband_2012} the quantity $L_\mathrm{QW,eff}^{\alpha}$ is defined 
%for a transition $\alpha$ between two subbands $\mu$, $\lambda$ by:
%$L_\mathrm{QW,eff}^{-1} = 
%					\frac{\hbar}{2m^{\ast}\omega_{\alpha}}
%					\int_{-\infty}^{+\infty}\left[ 
%					\Phi_{\mu}\partial_z\Phi_{\lambda}-\Phi_{\lambda}\partial_z\Phi_{\mu} 
%					\right]^2$
%where in our case, $\left[\lambda,\mu\right]=\left[1,2\right]$.
%}
The quantity $L_\mathrm{QW,eff}$ can be seen as an effective thickness of the quantum confined electron plasma 
and depends on the wave functions of the first and second subbands (see Ref. 41). 
We find $L_\mathrm{QW,eff}\approx25$~nm, smaller than the physical thickness of the QW (32~nm). 
For the following discussion, it is important to note that $\omega_P$ and hence $\tilde{\omega}$ depend 
on the charge density in a single quantum well only. 
The characteristic equation of the coupled intersubband plasmon-resonator system is written in the general case:
\begin{equation}
\left( \omega^2-\tilde{\omega}^{2} \right) \left( \omega^2-\omega_c^2 \right) = 
\Psi^2 f_{w} f_{12} \omega_P^2 \omega_c^2
\label{eq:dispersion}
\end{equation}
where $\omega_c$ is the resonator frequency, $f_{12}$ is the $1\rightarrow2$ transition oscillator strength ($f_{12}=0.96$ for an 
infinite QW), $f_{w}=N_{QW}L_\mathrm{QW,eff}/d$ describes the filling factor of the QWs inside the absorbing region with 
$N_{QW}$ charged quantum wells, and $\Psi^2$ describes the optical overlap of the resonator mode with 
the absorbing region, which is related to the effective mode volume $V_{\mathrm{eff}}$ as explained further. 
Note that the coefficient  $f_w$ quantifies the \emph{filling factor} of the QWs inside the absorbing region 
that takes into account the fact that the active dipoles are not homogeneously distributed in the semiconductor, 
but are only localized inside the QWs. 
Instead, the dimensionless overlap factor $\Psi^2$ represents the fraction of electromagnetic energy coupled into the 
$z$-component of the electric field and spatially overlapping with the semiconductor layers inside the capacitor. 
It is defined by:\cite{zanotto_analysis_2012} 

\begin{equation}
\Psi^2 =  \frac{\int_{AR}\frac{\varepsilon\varepsilon_0}{2}|E_z|^2}{U_e} 
\label{eq:Psi}
\end{equation}
where the energy stored in the vertical component of the electric field $E_z$ 
(the sole component of the field to couple to ISB transitions) is integrated over the absorbing region (AR) 
volume $V_{AR}=2W^2d$ and normalized by the total electric energy $U_e$ of the mode. 
The effective mode volume is then determined by the relation 
$V_{\mathrm{eff}}=V_{AR}/\Psi^2$. Since $\Psi^2 \leq 1$ this definition accounts for field leakage outside the 
capacitive parts of the resonator. 
Eq. \ref{eq:dispersion} provides the upper (UP) and lower (LP) polariton frequencies as a function of $\omega_c$. 
The minimum splitting between UP and LP is exactly the vacuum Rabi splitting $2\Omega_R$:

\begin{equation}
2\Omega_R = \sqrt{\Psi^2 f_w f_{12}} \omega_P = 
\sqrt{\frac{\Psi^2 f_{12} e^2 N_{QW}(n_1-n_2)}{\varepsilon \varepsilon_0m^{\ast}d}} 
\label{eq:Rabi}
\end{equation}

The Rabi splitting can also be expressed as  
$2\Omega_R = \sqrt{f_{12} e^2 /\varepsilon \varepsilon_0m^{\ast}} \times \sqrt{N_{QW}(N_1-N_2)/V_{\mathrm{eff}}}$, 
where $N_{1,2}$ is the total number of electrons populating the subband $1,2$. 
The Rabi splitting that is experimentally determined from spectroscopic studies, as described further, 
is thus directly linked to the effective mode volume $V_{\mathrm{eff}}$ which is an important quantity in 
nano-photonic systems.\cite{liu_fundamental_2016} 
However, the analysis of our experimental data is easier to perform with the help of Eq. \eqref{eq:Rabi}. 
Ultimately, two parameters govern the maximum coupling strength: 
the overlap factor $\Psi^2$ and the number of available dipoles in the microcavity volume $N_{QW}(n_1-n_2)$. 
The strength of the coupling can thus be controlled by tuning the population difference through 
the change of the temperature of the sample.

In order to asses the parameters $N_{QW}$ and $\Psi^2$ in our LC resonators, we compare the spectroscopic features 
of the ultra-strong coupling regime in the LC resonators with square patch double-metal microcavities. 
Such double-metal cavities sustain a resonance at $\lambda=2n_{eff}s$, 
where $s$ is the size of the patch, and $n_{eff}=3.9$ the effective index of the confined mode.\cite{todorov_optical_2010} 
They have been shown to reach the ultra-strong coupling regime with similar absorbing regions with $\Psi^2 \approx 1$, 
however with a much larger effective mode volume.
\cite{todorov_ultrastrong_2010,todorov_polaritonic_2012} 
They will serve as a reference to our current LC samples.\cite{feuillet-palma_extremely_2012}

\begin{figure*}[!ht]
\centering
\includegraphics[width=\linewidth]{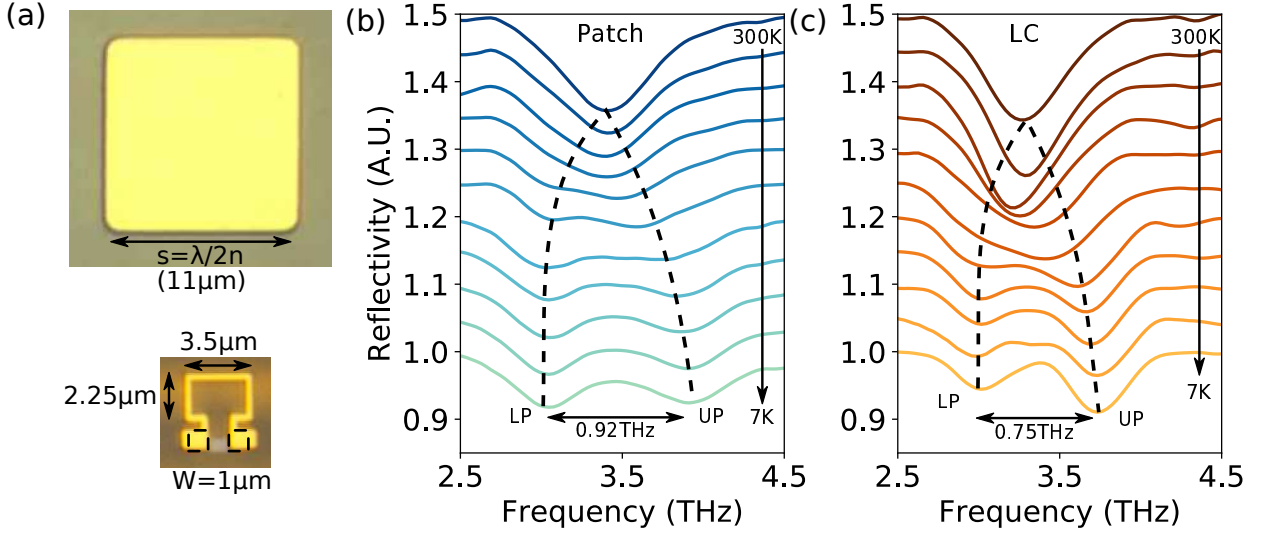}
\caption{
(a) Optical microscope image of a patch microcavity (top) and a LC resonator (bottom) to scale. 
The patch acts as a double-metal Fabry-Perot cavity for TM$_0$ guided modes, where the first resonant frequency is given 
by $\lambda=2ns$. In the LC resonator, we have highlighted in dashed squares the capacitor area, 
evidencing the dramatic reduction of the electric mode volume. 
(b) Reflectivity of the patch microcavity array as a function of temperature. The resonator mode visible at room temperature 
splits into upper (UP) and lower (LP) polariton modes, with a separation of $0.92$~THz. 
(c) Same experiment for the LC metamaterial, resulting in a separation of $0.75$~THz.
The SiN baseline has been removed for clarity. The dashed lines are guides for the eye.
}
\label{fig:TSweep}
\end{figure*}

A comparison between a LC resonator and a patch cavity with the same absorbing region is shown in Fig.~\ref{fig:TSweep}. 
Optical microscope images of the two types of resonators, which have identical resonant frequencies (3.5 THz) are shown 
in Fig.~\ref{fig:TSweep}~(a) (to scale). While the size of the patch cavity is set by the diffraction limit, 
clearly the lateral confinement of the electric field in the capacitors of the LC resonator is well below that limit.  
For spectroscopic studies at cryogenic temperatures, the samples are mounted on the cold finger of a liquid helium 
continuous flow cryostat, and probed in a reflectivity experiment. 
We report in Fig.~\ref{fig:TSweep}~(b) (resp. (c)) the reflectivity of the patch cavities array (resp. LC resonator) 
tuned near the intersubband plasmon resonance as a function of temperature, ranging from 300~K to 7~K. 
In the case of the LC resonator, the baseline induced by the SiN has been removed for clarity. 
In both cases the room temperature spectra show a single resonance around 3.5~THz 
as the population difference between the first two subbands is zero ($n_1 \approx n_2$), 
owing to the thermal electron distribution. 
The matter excitation thus vanishes, and one only sees the LC and the patch cavity resonances. 
This single resonance splits when decreasing the temperature as the population difference $n_1-n_2$ increases, 
and the maximum separation is obtained at low temperature (7K), reaching 0.92~THz for the patch cavities and 
0.75~THz for the LC resonators. 

\begin{figure*}[!ht]
\centering
\includegraphics[width=\linewidth]{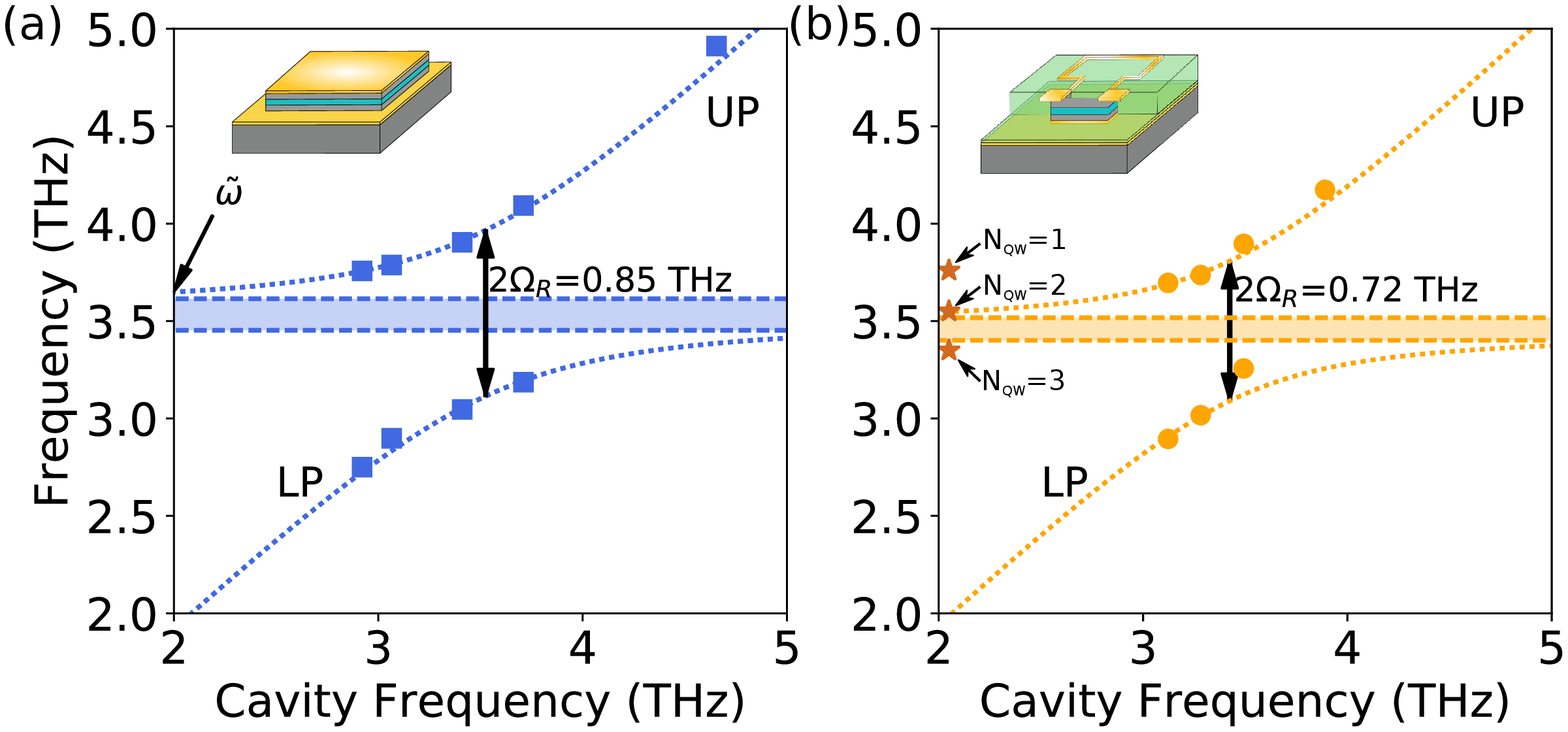}
\caption{
Dispersion of the LP and UP modes as a function of cavity frequency for patch microcavities (a) or 
LC resonators (b). Dotted lines show the polariton dispersion relation obtained from eq.~\eqref{eq:dispersion}. 
A polaritonic gap appears between the UP and LP branches (dashes and filled area).  
Stars on panel (b) indicate the calculated intersubband plasmon frequency in the case where $N_{QW}=1,2,3$ 
(see main text and Supplementary Information).
}
\label{fig:USCLCs}
\end{figure*}

Tuning the LC and patch resonant frequencies allows us to more precisely map the dispersion relation 
given in equation \eqref{eq:dispersion}. 
In Fig.~\ref{fig:USCLCs} the position of the UP and LP modes at T=7~K are plotted 
as a function of the cavity frequency, obtained by modifying either the size $s$ of the patch cavities, 
or the inductance of the LC resonators through the internal perimeter P$_{int}$. 
We observe a clear anticrossing and the opening of a polaritonic gap between the two polariton branches. 
By fitting the data with equation \eqref{eq:dispersion}, we can extract the Rabi splittings 
$2\Omega_{R-LC}=0.72$~THz and $2\Omega_{R-patchs}=0.85$~THz, representing respectively 21\% and 24\% 
of the intersubband plasmon frequency $\tilde{\omega}$ determined from the fit of the dispersion relation. 
The slight reduction of the Rabi splitting for LC resonators can be due to both lateral depletion of the QW, 
as well as to the overlap factor $\Psi^2$ that is less than unity.

We now want to determine separately the total charge and the overlap factor in our systems. 
We start by deducing the total charge left in the 5~QW patch cavity sample, by comparing it to a reference sample 
having exactly the same QWs, but repeated 25 times. 
The 25 QWs sample is also processed into patch cavities (Refs. \citenum{todorov_ultrastrong_2010,todorov_polaritonic_2012}), 
and we can safely assume that $\Psi_{ref}^2=\Psi^2=1$. 
The data for that sample are provided in the Supplementary Materials, along with a detailed derivation of the method to extract 
the electronic population. We deduce that 21 QWs are populated, with an equivalent doping of $1.37\times10^{11}$cm$^{-2}$. 
Using equation \eqref{eq:Rabi}, (see also Ref. \citenum{todorov_ultrastrong_2010,todorov_polaritonic_2012}) 
a proportionality rule yields the total surface charge density at low temperature: 
\begin{equation}
\left(\frac{\Omega_R}{\Omega_{R-ref}}\right)^2  
						% =  \frac{\Psi^2 f_wf_{12}}{\Psi_{ref}^2 f_{w-ref}f_{12-ref}}   %%%%%% commented out
							= \frac{\Psi^2 f_{12}N_{QW} n_1}{\Psi_{ref}^2 f_{12-ref}N_{QW-ref} n_{1-ref}}
												\frac{d_{ref}}{d} \label{eq:OmegaR_ratio}
\end{equation} 
In that case $\Psi^2=\Psi_{ref}^2=1$, and we find an equivalent total charge density of $N_{QW} n_1=2 n_1$, 
meaning that only $2/5$ of the total charge is left is the QWs, 
confirming the importance of the depletion effects at the  metal-semiconductor interfaces. 
Furthermore, for both samples we observe the same matter excitation frequency $\tilde{\omega}$, 
which means the surface charge density per quantum well is the same. 
We therefore conclude that we have $N_{QW}=$2 charged quantum wells in the 5QW absorbing region. 
A more detailed analysis is given in the Supplementary Information. 

Having determined the total charge in the case of the patch cavities, we use the same proportionality rule 
to compare the 5QW patch and LC resonators. We first assume that we have the same total charge in both samples. 
Then, according to equation \eqref{eq:OmegaR_ratio} we can derive the geometric overlap factor $\Psi^2$. 
Comparing the data from Fig.~\ref{fig:USCLCs} we derive a confinement factor $\Psi^2=0.7$. 
A more careful comparison of Fig.~\ref{fig:USCLCs}~(a) and (b) shows that the fits of the dispersion relations 
yields a plasmon frequency $\tilde{\omega}$ at a slightly lower frequency for the LC resonators ($\tilde{\omega}=$3.55~THz) 
than the one of the patch cavities ($\tilde{\omega}=$3.65~THz). 
Since the plasmon frequency is related to the plasma frequency through the formula  
$\tilde{\omega}=\sqrt{\omega_{12}^2+\omega_{P}^2}$, 
we conclude that the red shift is due to a lower plasma frequency of the LC as compared to the patch cavity. 
This is due to a further lateral depletion of the QWs originating from surface traps at the etched surface 
of the absorbing region, as already reported in the case of etched pillar or nanowire structures.
\cite{dietze_ultrastrong_2013,amanti_electrically_2013,lahnemann_near-infrared_2017} 
From our data, we infer a plasma frequency for the LC resonators that is 7.6\% lower than the one for the patch cavities. 
Correcting for this effect yields an overlap factor $\Psi^2=0.79$. 
This value is slightly larger than the one predicted in numerical simulations, $\Psi_{LC}^2=0.64$. 
The effective mode volume of the LC resonators is thus determined to be 
$V_{\mathrm{eff}}=1.2\times10^{-6}\lambda_0^3$, almost two orders of magnitude smaller than the patch cavities. 
The low value of the effective mode volume is a striking feature of our resonator, 
since achieving an effective volume very close to the physical volume of 
the semiconductor absorbing region ($9.2\times10^{-7}\lambda_0^3$) represents a critical trade-off in double-metal geometries, 
owing to the leakage of the electric energy in fringing fields. 
Square or wire patch cavities indeed lead to $\Psi^2$ factors close to unity at the expense of a large 
mode volume, while other systems report very small mode volume, sacrificing the overlap factor down to a 
few percents. \cite{malerba_towards_2016,paulillo_room_2016}
We can also estimate the intersubband plasmon frequency $\tilde{\omega}$ in the case where one or three QWs would be 
populated. The detail of the calculation is given in Supplementary Information, and the results are shown in 
stars in Fig.~\ref{fig:USCLCs}~(b). We can see that the intersubband plasmon frequencies obtained in the two cases strongly 
differ from our measurement, confirming the analysis presented above. 

From the knowledge of the equivalent surface charge density we can deduce the total number of electrons 
participating in the coupling with the cavity and resonator modes. 
The calculation yields $N_{e-patchs} = 3.3 \times 10^5 $~e$^{-}$/patch and  
$N_{e-LC} = 2.4 \times 10^3 $~e$^{-}$/capacitor. Our newly developed LC resonators thus allow us to greatly decrease the 
number of electrons involved in  the coupling while maintaining a large vacuum Rabi splitting, making a step towards 
the few electrons regime beyond previous results on double-metal cavities.
\cite{feuillet-palma_extremely_2012,paulillo_room_2016,malerba_towards_2016}

%\section{Increasing doping}

\begin{figure*}[!ht]
\centering
\includegraphics[width=\linewidth]{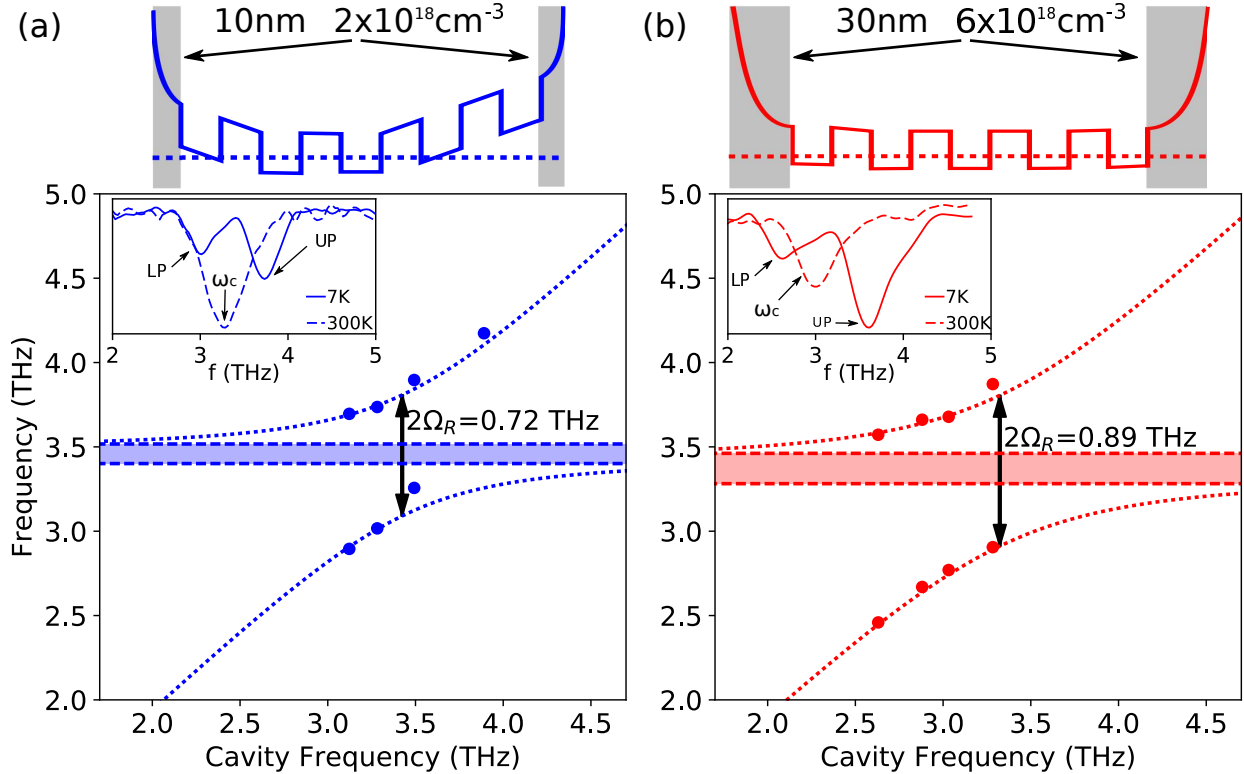}
\caption{
Polariton dispersion relations for sample with different doping densities on each side of the absorbing region. 
(a) Reproduction of the dispersion relation in Fig.~\ref{fig:USCLCs}~(b). 
The outer layers are doped at $2\times 10^{18}$cm$^{-3}$ over 10~nm.
(b) Polariton dispersion relation for LC resonators with outer layers doped at $6\times10^{18}$cm$^{-3}$ over 30~nm. 
The top of each panels sketches the conduction band profile, evidencing the depletion due to the band 
bending at the metal-semiconductor interfaces. Insets: reflectivity spectra at room temperature and 7~K 
for both samples, evidencing the cavity frequency, and the LP and UP resonances. 
}
\label{fig:USCDop}
\end{figure*}

In order to improve further the strength of the light-matter coupling, we designed a second sample 
with an identical absorbing region, but with an increased thickness of the cavity (320~nm) and and increased doping density in 
the outermost GaAs layers ($6\times10^{18}$~cm$^{-3}$). 
After processing into patch cavities and LC metamaterials, we perform the same experiments as described above 
(see Supplementary materials for details). 
Our results show that in the present sample, all 5 QWs are populated. 
This means that thanks to the increased charge density in the interface layers, 
the populated QWs participating in the optical absorption occupy a larger fraction of the total cavity volume, 
increasing the filling factor $f_w$. 
However, we deduce a lower value of the optical overlap factor, $\Psi^2$=0.56, meaning a slightly higher effective volume 
than that of the previous LC sample. 
This could be explained by a partial screening of the penetration of the electric field inside the capacitors, 
owing to the higher number of free carriers in the contact layers. 
Nevertheless, as shown in Fig.~\ref{fig:USCDop}, the Rabi splitting $2\Omega_R$ is increased to 0.89~THz in the new LC 
sample, reaching values of 0.27 of the ISB plasmon frequency. 
The total electron number is $N_e=6\times10^3$e$^-$/capacitor. 
Remarkably, this value is one of the lowest achieved so far using ISB transitions coupled to metamaterials, 
while retaining a large $\Omega_R/\omega_c$ ratio. 

%\section{Conclusion}

In conclusion, we demonstrate a deeply subwavelength confinement of electromagnetic energy with an 
effective mode volume $V_{\mathrm{eff}}=10^{-6}\lambda_0^3$ and a reduction of the number of interacting electrons, 
down to a few thousands while keeping $2\Omega_R/\tilde{\omega}=0.27$. 
The electron number is an order of magnitude larger than that recently reported in systems 
exploring the coupling between Landau level transitions and planar metamaterials.\cite{keller_few-electron_2017} 
However, our structures will allow further reduction of the number of electrons while maintaining a large ratio $\Omega_R/\tilde{\omega}$. 
This can be achieved by reducing the size of the capacitive elements to a few hundreds of nanometers in our LC 
resonators. 
For instance, by maintaining a surface equivalent doping of $\approx2\times10^{11}$cm$^{-2}$ 
and working with a single quantum well, we could achieve ultra-strong light-matter coupling with only 10 electrons in 
100~nm wide capacitors. Devices with such small capacitive elements have already been fabricated in SiN-based 
LC resonators.\cite{mottaghizadeh_nanoscale_2017}

An important asset of our device architectures is that they naturally provide the possibility to implement 
DC current input/output from the semiconductor region.
Indeed, the top and bottom metallic parts are direct current uncoupled, and contacts can be taken directly 
on the inductive parts without hindering the sub-wavelength confinement in the capacitors. 
Such device architectures will allow us to study new effects related to the ultra-strong coupling, beyond the spectroscopic studies 
performed so far,\cite{keller_few-electron_2017} such as polariton-assisted fermionic transport.\cite{feist_extraordinary_2015,orgiu_conductivity_2015,bruhat_cavity_2016,stockklauser_strong_2017,paravicini-bagliani_magneto-transport_2018,hagenmuller_intrinsic_2018}  
The reduced number of electrons leads to a reduced number of dark states, 
optimizing the coupling between the collective electronic excitation and the electronic current.\cite{de_liberato_quantum_2009} 
Furthermore, in the weak coupling regime, these device architectures can be very beneficial for 
ultra-low dark current ISB detectors.\cite{palaferri_ultra-subwavelength_2016}


\begin{acknowledgements}

The authors acknowledge the help of the technical staff from the cleanroom facility of Université Paris Diderot.  
We thank Li Chen for the wafer bonding of the samples used to fabricate patch antennas. 
This work was supported by the French National Research Agency under the contract ANR-16-CE24-0020. 
We acknowledge support from the EPSRC (UK) programme grant HyperTerahertz EP/P021859/1.  
EHL acknowledges support from the Royal Society and the Wolfson Foundation.

\end{acknowledgements}

\bibliography{Bib_USC-THz-LC.bib}

\begin{thebibliography}{53}%
\makeatletter
\providecommand \@ifxundefined [1]{%
 \@ifx{#1\undefined}
}%
\providecommand \@ifnum [1]{%
 \ifnum #1\expandafter \@firstoftwo
 \else \expandafter \@secondoftwo
 \fi
}%
\providecommand \@ifx [1]{%
 \ifx #1\expandafter \@firstoftwo
 \else \expandafter \@secondoftwo
 \fi
}%
\providecommand \natexlab [1]{#1}%
\providecommand \enquote  [1]{``#1''}%
\providecommand \bibnamefont  [1]{#1}%
\providecommand \bibfnamefont [1]{#1}%
\providecommand \citenamefont [1]{#1}%
\providecommand \href@noop [0]{\@secondoftwo}%
\providecommand \href [0]{\begingroup \@sanitize@url \@href}%
\providecommand \@href[1]{\@@startlink{#1}\@@href}%
\providecommand \@@href[1]{\endgroup#1\@@endlink}%
\providecommand \@sanitize@url [0]{\catcode `\\12\catcode `\$12\catcode
  `\&12\catcode `\#12\catcode `\^12\catcode `\_12\catcode `\%12\relax}%
\providecommand \@@startlink[1]{}%
\providecommand \@@endlink[0]{}%
\providecommand \url  [0]{\begingroup\@sanitize@url \@url }%
\providecommand \@url [1]{\endgroup\@href {#1}{\urlprefix }}%
\providecommand \urlprefix  [0]{URL }%
\providecommand \Eprint [0]{\href }%
\providecommand \doibase [0]{http://dx.doi.org/}%
\providecommand \selectlanguage [0]{\@gobble}%
\providecommand \bibinfo  [0]{\@secondoftwo}%
\providecommand \bibfield  [0]{\@secondoftwo}%
\providecommand \translation [1]{[#1]}%
\providecommand \BibitemOpen [0]{}%
\providecommand \bibitemStop [0]{}%
\providecommand \bibitemNoStop [0]{.\EOS\space}%
\providecommand \EOS [0]{\spacefactor3000\relax}%
\providecommand \BibitemShut  [1]{\csname bibitem#1\endcsname}%
\let\auto@bib@innerbib\@empty
%</preamble>
\bibitem [{\citenamefont {Shelby}(2001)}]{shelby_experimental_2001}%
  \BibitemOpen
  \bibfield  {author} {\bibinfo {author} {\bibfnamefont {R.~A.}\ \bibnamefont
  {Shelby}},\ }\href {\doibase 10.1126/science.1058847} {\bibfield  {journal}
  {\bibinfo  {journal} {Science}\ }\textbf {\bibinfo {volume} {292}},\ \bibinfo
  {pages} {77} (\bibinfo {year} {2001})}\BibitemShut {NoStop}%
\bibitem [{\citenamefont {Pendry}\ \emph {et~al.}(1999)\citenamefont {Pendry},
  \citenamefont {Holden}, \citenamefont {Robbins},\ and\ \citenamefont
  {Stewart}}]{pendry_magnetism_1999}%
  \BibitemOpen
  \bibfield  {author} {\bibinfo {author} {\bibfnamefont {J.}~\bibnamefont
  {Pendry}}, \bibinfo {author} {\bibfnamefont {A.}~\bibnamefont {Holden}},
  \bibinfo {author} {\bibfnamefont {D.}~\bibnamefont {Robbins}}, \ and\
  \bibinfo {author} {\bibfnamefont {W.}~\bibnamefont {Stewart}},\ }\href
  {\doibase 10.1109/22.798002} {\bibfield  {journal} {\bibinfo  {journal} {IEEE
  Transactions on Microwave Theory and Techniques}\ }\textbf {\bibinfo {volume}
  {47}},\ \bibinfo {pages} {2075} (\bibinfo {year} {Nov./1999})}\BibitemShut
  {NoStop}%
\bibitem [{\citenamefont {Cai}\ and\ \citenamefont {{\v
  S}alaev}(2010)}]{cai_optical_2010}%
  \BibitemOpen
  \bibfield  {author} {\bibinfo {author} {\bibfnamefont {W.}~\bibnamefont
  {Cai}}\ and\ \bibinfo {author} {\bibfnamefont {V.~M.}\ \bibnamefont {{\v
  S}alaev}},\ }\href@noop {} {{\selectlanguage {english}\emph {\bibinfo {title}
  {Optical Metamaterials: Fundamentals and Applications}}}}\ (\bibinfo
  {publisher} {{Springer}},\ \bibinfo {address} {New York, NY},\ \bibinfo
  {year} {2010})\ \bibinfo {note} {oCLC: 837215261}\BibitemShut {NoStop}%
\bibitem [{\citenamefont {Wang}\ \emph {et~al.}(2017)\citenamefont {Wang},
  \citenamefont {Wang}, \citenamefont {Li}, \citenamefont {Chen}, \citenamefont
  {Wang}, \citenamefont {Dai}, \citenamefont {Oulton},\ and\ \citenamefont
  {Ma}}]{wang_unusual_2017}%
  \BibitemOpen
  \bibfield  {author} {\bibinfo {author} {\bibfnamefont {S.}~\bibnamefont
  {Wang}}, \bibinfo {author} {\bibfnamefont {X.-Y.}\ \bibnamefont {Wang}},
  \bibinfo {author} {\bibfnamefont {B.}~\bibnamefont {Li}}, \bibinfo {author}
  {\bibfnamefont {H.-Z.}\ \bibnamefont {Chen}}, \bibinfo {author}
  {\bibfnamefont {Y.-L.}\ \bibnamefont {Wang}}, \bibinfo {author}
  {\bibfnamefont {L.}~\bibnamefont {Dai}}, \bibinfo {author} {\bibfnamefont
  {R.~F.}\ \bibnamefont {Oulton}}, \ and\ \bibinfo {author} {\bibfnamefont
  {R.-M.}\ \bibnamefont {Ma}},\ }\href {\doibase 10.1038/s41467-017-01662-6}
  {\bibfield  {journal} {\bibinfo  {journal} {Nature Communications}\ }\textbf
  {\bibinfo {volume} {8}},\ \bibinfo {pages} {1889} (\bibinfo {year}
  {2017})}\BibitemShut {NoStop}%
\bibitem [{\citenamefont {Chen}\ \emph {et~al.}(2012)\citenamefont {Chen},
  \citenamefont {Li},\ and\ \citenamefont {Sun}}]{chen_metamaterials_2012}%
  \BibitemOpen
  \bibfield  {author} {\bibinfo {author} {\bibfnamefont {T.}~\bibnamefont
  {Chen}}, \bibinfo {author} {\bibfnamefont {S.}~\bibnamefont {Li}}, \ and\
  \bibinfo {author} {\bibfnamefont {H.}~\bibnamefont {Sun}},\ }\href {\doibase
  10.3390/s120302742} {\bibfield  {journal} {\bibinfo  {journal} {Sensors}\
  }\textbf {\bibinfo {volume} {12}},\ \bibinfo {pages} {2742} (\bibinfo {year}
  {2012})}\BibitemShut {NoStop}%
\bibitem [{\citenamefont {Belacel}\ \emph {et~al.}(2017)\citenamefont
  {Belacel}, \citenamefont {Todorov}, \citenamefont {Barbieri}, \citenamefont
  {Gacemi}, \citenamefont {Favero},\ and\ \citenamefont
  {Sirtori}}]{belacel_optomechanical_2017}%
  \BibitemOpen
  \bibfield  {author} {\bibinfo {author} {\bibfnamefont {C.}~\bibnamefont
  {Belacel}}, \bibinfo {author} {\bibfnamefont {Y.}~\bibnamefont {Todorov}},
  \bibinfo {author} {\bibfnamefont {S.}~\bibnamefont {Barbieri}}, \bibinfo
  {author} {\bibfnamefont {D.}~\bibnamefont {Gacemi}}, \bibinfo {author}
  {\bibfnamefont {I.}~\bibnamefont {Favero}}, \ and\ \bibinfo {author}
  {\bibfnamefont {C.}~\bibnamefont {Sirtori}},\ }\href {\doibase
  10.1038/s41467-017-01840-6} {\bibfield  {journal} {\bibinfo  {journal}
  {Nature Communications}\ }\textbf {\bibinfo {volume} {8}},\ \bibinfo {pages}
  {1578} (\bibinfo {year} {2017})}\BibitemShut {NoStop}%
\bibitem [{\citenamefont {Alves}\ \emph {et~al.}(2018)\citenamefont {Alves},
  \citenamefont {Pimental}, \citenamefont {Grbovic},\ and\ \citenamefont
  {Karunasiri}}]{alves_mems_2018}%
  \BibitemOpen
  \bibfield  {author} {\bibinfo {author} {\bibfnamefont {F.}~\bibnamefont
  {Alves}}, \bibinfo {author} {\bibfnamefont {L.}~\bibnamefont {Pimental}},
  \bibinfo {author} {\bibfnamefont {D.}~\bibnamefont {Grbovic}}, \ and\
  \bibinfo {author} {\bibfnamefont {G.}~\bibnamefont {Karunasiri}},\ }\href
  {\doibase 10.1038/s41598-018-30858-z} {\bibfield  {journal} {\bibinfo
  {journal} {Scientific Reports}\ }\textbf {\bibinfo {volume} {8}},\ \bibinfo
  {pages} {12466} (\bibinfo {year} {2018})}\BibitemShut {NoStop}%
\bibitem [{\citenamefont {Shrekenhamer}\ \emph {et~al.}(2012)\citenamefont
  {Shrekenhamer}, \citenamefont {Xu}, \citenamefont {Venkatesh}, \citenamefont
  {Schurig}, \citenamefont {Sonkusale},\ and\ \citenamefont
  {Padilla}}]{shrekenhamer_experimental_2012}%
  \BibitemOpen
  \bibfield  {author} {\bibinfo {author} {\bibfnamefont {D.}~\bibnamefont
  {Shrekenhamer}}, \bibinfo {author} {\bibfnamefont {W.}~\bibnamefont {Xu}},
  \bibinfo {author} {\bibfnamefont {S.}~\bibnamefont {Venkatesh}}, \bibinfo
  {author} {\bibfnamefont {D.}~\bibnamefont {Schurig}}, \bibinfo {author}
  {\bibfnamefont {S.}~\bibnamefont {Sonkusale}}, \ and\ \bibinfo {author}
  {\bibfnamefont {W.~J.}\ \bibnamefont {Padilla}},\ }\href {\doibase
  10.1103/PhysRevLett.109.177401} {\bibfield  {journal} {\bibinfo  {journal}
  {Physical Review Letters}\ }\textbf {\bibinfo {volume} {109}},\ \bibinfo
  {pages} {177401} (\bibinfo {year} {2012})}\BibitemShut {NoStop}%
\bibitem [{\citenamefont {Viti}\ \emph {et~al.}(2014)\citenamefont {Viti},
  \citenamefont {Coquillat}, \citenamefont {Ercolani}, \citenamefont {Sorba},
  \citenamefont {Knap},\ and\ \citenamefont {Vitiello}}]{viti_nanowire_2014}%
  \BibitemOpen
  \bibfield  {author} {\bibinfo {author} {\bibfnamefont {L.}~\bibnamefont
  {Viti}}, \bibinfo {author} {\bibfnamefont {D.}~\bibnamefont {Coquillat}},
  \bibinfo {author} {\bibfnamefont {D.}~\bibnamefont {Ercolani}}, \bibinfo
  {author} {\bibfnamefont {L.}~\bibnamefont {Sorba}}, \bibinfo {author}
  {\bibfnamefont {W.}~\bibnamefont {Knap}}, \ and\ \bibinfo {author}
  {\bibfnamefont {M.~S.}\ \bibnamefont {Vitiello}},\ }\href {\doibase
  10.1364/OE.22.008996} {\bibfield  {journal} {\bibinfo  {journal} {Optics
  Express}\ }\textbf {\bibinfo {volume} {22}},\ \bibinfo {pages} {8996}
  (\bibinfo {year} {2014})}\BibitemShut {NoStop}%
\bibitem [{\citenamefont {Luxmoore}\ \emph {et~al.}(2016)\citenamefont
  {Luxmoore}, \citenamefont {Liu}, \citenamefont {Li}, \citenamefont {Faist},\
  and\ \citenamefont {Nash}}]{luxmoore_graphenemetamaterial_2016}%
  \BibitemOpen
  \bibfield  {author} {\bibinfo {author} {\bibfnamefont {I.~J.}\ \bibnamefont
  {Luxmoore}}, \bibinfo {author} {\bibfnamefont {P.~Q.}\ \bibnamefont {Liu}},
  \bibinfo {author} {\bibfnamefont {P.}~\bibnamefont {Li}}, \bibinfo {author}
  {\bibfnamefont {J.}~\bibnamefont {Faist}}, \ and\ \bibinfo {author}
  {\bibfnamefont {G.~R.}\ \bibnamefont {Nash}},\ }\href {\doibase
  10.1021/acsphotonics.6b00226} {\bibfield  {journal} {\bibinfo  {journal} {ACS
  Photonics}\ }\textbf {\bibinfo {volume} {3}},\ \bibinfo {pages} {936}
  (\bibinfo {year} {2016})}\BibitemShut {NoStop}%
\bibitem [{\citenamefont {Palaferri}\ \emph {et~al.}(2016)\citenamefont
  {Palaferri}, \citenamefont {Todorov}, \citenamefont {Mottaghizadeh},
  \citenamefont {Frucci}, \citenamefont {Biasiol},\ and\ \citenamefont
  {Sirtori}}]{palaferri_ultra-subwavelength_2016}%
  \BibitemOpen
  \bibfield  {author} {\bibinfo {author} {\bibfnamefont {D.}~\bibnamefont
  {Palaferri}}, \bibinfo {author} {\bibfnamefont {Y.}~\bibnamefont {Todorov}},
  \bibinfo {author} {\bibfnamefont {A.}~\bibnamefont {Mottaghizadeh}}, \bibinfo
  {author} {\bibfnamefont {G.}~\bibnamefont {Frucci}}, \bibinfo {author}
  {\bibfnamefont {G.}~\bibnamefont {Biasiol}}, \ and\ \bibinfo {author}
  {\bibfnamefont {C.}~\bibnamefont {Sirtori}},\ }\href {\doibase
  10.1088/1367-2630/18/11/113016} {\bibfield  {journal} {\bibinfo  {journal}
  {New Journal of Physics}\ }\textbf {\bibinfo {volume} {18}},\ \bibinfo
  {pages} {113016} (\bibinfo {year} {2016})}\BibitemShut {NoStop}%
\bibitem [{\citenamefont {Palaferri}\ \emph {et~al.}(2018)\citenamefont
  {Palaferri}, \citenamefont {Todorov}, \citenamefont {Bigioli}, \citenamefont
  {Mottaghizadeh}, \citenamefont {Gacemi}, \citenamefont {Calabrese},
  \citenamefont {Vasanelli}, \citenamefont {Li}, \citenamefont {Davies},
  \citenamefont {Linfield}, \citenamefont {Kapsalidis}, \citenamefont {Beck},
  \citenamefont {Faist},\ and\ \citenamefont
  {Sirtori}}]{palaferri_room-temperature_2018}%
  \BibitemOpen
  \bibfield  {author} {\bibinfo {author} {\bibfnamefont {D.}~\bibnamefont
  {Palaferri}}, \bibinfo {author} {\bibfnamefont {Y.}~\bibnamefont {Todorov}},
  \bibinfo {author} {\bibfnamefont {A.}~\bibnamefont {Bigioli}}, \bibinfo
  {author} {\bibfnamefont {A.}~\bibnamefont {Mottaghizadeh}}, \bibinfo {author}
  {\bibfnamefont {D.}~\bibnamefont {Gacemi}}, \bibinfo {author} {\bibfnamefont
  {A.}~\bibnamefont {Calabrese}}, \bibinfo {author} {\bibfnamefont
  {A.}~\bibnamefont {Vasanelli}}, \bibinfo {author} {\bibfnamefont
  {L.}~\bibnamefont {Li}}, \bibinfo {author} {\bibfnamefont {A.~G.}\
  \bibnamefont {Davies}}, \bibinfo {author} {\bibfnamefont {E.~H.}\
  \bibnamefont {Linfield}}, \bibinfo {author} {\bibfnamefont {F.}~\bibnamefont
  {Kapsalidis}}, \bibinfo {author} {\bibfnamefont {M.}~\bibnamefont {Beck}},
  \bibinfo {author} {\bibfnamefont {J.}~\bibnamefont {Faist}}, \ and\ \bibinfo
  {author} {\bibfnamefont {C.}~\bibnamefont {Sirtori}},\ }\href {\doibase
  10.1038/nature25790} {\bibfield  {journal} {\bibinfo  {journal} {Nature}\
  }\textbf {\bibinfo {volume} {556}},\ \bibinfo {pages} {85} (\bibinfo {year}
  {2018})}\BibitemShut {NoStop}%
\bibitem [{\citenamefont {Louisell}(1990)}]{louisell_quantum_1990}%
  \BibitemOpen
  \bibfield  {author} {\bibinfo {author} {\bibfnamefont {W.~H.}\ \bibnamefont
  {Louisell}},\ }\href@noop {} {{\selectlanguage {english}\emph {\bibinfo
  {title} {Quantum Statistical Properties of Radiation}}}},\ Wiley Classics
  Library\ (\bibinfo  {publisher} {{Wiley}},\ \bibinfo {address} {New York},\
  \bibinfo {year} {1990})\ \bibinfo {note} {oCLC: 833244245}\BibitemShut
  {NoStop}%
\bibitem [{\citenamefont {Purcell}\ \emph {et~al.}(1946)\citenamefont
  {Purcell}, \citenamefont {Torrey},\ and\ \citenamefont
  {Pound}}]{purcell_resonance_1946}%
  \BibitemOpen
  \bibfield  {author} {\bibinfo {author} {\bibfnamefont {E.~M.}\ \bibnamefont
  {Purcell}}, \bibinfo {author} {\bibfnamefont {H.~C.}\ \bibnamefont {Torrey}},
  \ and\ \bibinfo {author} {\bibfnamefont {R.~V.}\ \bibnamefont {Pound}},\
  }\href {\doibase 10.1103/PhysRev.69.37} {\bibfield  {journal} {\bibinfo
  {journal} {Physical Review}\ }\textbf {\bibinfo {volume} {69}},\ \bibinfo
  {pages} {37} (\bibinfo {year} {1946})}\BibitemShut {NoStop}%
\bibitem [{\citenamefont {Yamamoto}\ \emph {et~al.}(2000)\citenamefont
  {Yamamoto}, \citenamefont {Tassone},\ and\ \citenamefont
  {Cao}}]{yamamoto_semiconductor_2000}%
  \BibitemOpen
  \bibfield  {author} {\bibinfo {author} {\bibfnamefont {Y.}~\bibnamefont
  {Yamamoto}}, \bibinfo {author} {\bibfnamefont {F.}~\bibnamefont {Tassone}}, \
  and\ \bibinfo {author} {\bibfnamefont {H.}~\bibnamefont {Cao}},\ }\href@noop
  {} {\emph {\bibinfo {title} {Semiconductor {{Cavity Quantum
  Electrodynamics}}}}},\ \bibinfo {edition} {springer}\ ed.\ (\bibinfo {year}
  {2000})\BibitemShut {NoStop}%
\bibitem [{\citenamefont {Baranov}\ \emph {et~al.}(2018)\citenamefont
  {Baranov}, \citenamefont {Wers\"all}, \citenamefont {Cuadra}, \citenamefont
  {Antosiewicz},\ and\ \citenamefont {Shegai}}]{baranov_novel_2018}%
  \BibitemOpen
  \bibfield  {author} {\bibinfo {author} {\bibfnamefont {D.~G.}\ \bibnamefont
  {Baranov}}, \bibinfo {author} {\bibfnamefont {M.}~\bibnamefont {Wers\"all}},
  \bibinfo {author} {\bibfnamefont {J.}~\bibnamefont {Cuadra}}, \bibinfo
  {author} {\bibfnamefont {T.~J.}\ \bibnamefont {Antosiewicz}}, \ and\ \bibinfo
  {author} {\bibfnamefont {T.}~\bibnamefont {Shegai}},\ }\href {\doibase
  10.1021/acsphotonics.7b00674} {\bibfield  {journal} {\bibinfo  {journal} {ACS
  Photonics}\ }\textbf {\bibinfo {volume} {5}},\ \bibinfo {pages} {24}
  (\bibinfo {year} {2018})}\BibitemShut {NoStop}%
\bibitem [{\citenamefont {Frisk~Kockum}\ \emph {et~al.}(2019)\citenamefont
  {Frisk~Kockum}, \citenamefont {Miranowicz}, \citenamefont {De~Liberato},
  \citenamefont {Savasta},\ and\ \citenamefont
  {Nori}}]{frisk_kockum_ultrastrong_2019}%
  \BibitemOpen
  \bibfield  {author} {\bibinfo {author} {\bibfnamefont {A.}~\bibnamefont
  {Frisk~Kockum}}, \bibinfo {author} {\bibfnamefont {A.}~\bibnamefont
  {Miranowicz}}, \bibinfo {author} {\bibfnamefont {S.}~\bibnamefont
  {De~Liberato}}, \bibinfo {author} {\bibfnamefont {S.}~\bibnamefont
  {Savasta}}, \ and\ \bibinfo {author} {\bibfnamefont {F.}~\bibnamefont
  {Nori}},\ }\href {\doibase 10.1038/s42254-018-0006-2} {\bibfield  {journal}
  {\bibinfo  {journal} {Nature Reviews Physics}\ }\textbf {\bibinfo {volume}
  {1}},\ \bibinfo {pages} {19} (\bibinfo {year} {2019})}\BibitemShut {NoStop}%
\bibitem [{\citenamefont {{Forn-D\'iaz}}\ \emph {et~al.}(2018)\citenamefont
  {{Forn-D\'iaz}}, \citenamefont {Lamata}, \citenamefont {Rico}, \citenamefont
  {Kono},\ and\ \citenamefont {Solano}}]{forn-diaz_ultrastrong_2018}%
  \BibitemOpen
  \bibfield  {author} {\bibinfo {author} {\bibfnamefont {P.}~\bibnamefont
  {{Forn-D\'iaz}}}, \bibinfo {author} {\bibfnamefont {L.}~\bibnamefont
  {Lamata}}, \bibinfo {author} {\bibfnamefont {E.}~\bibnamefont {Rico}},
  \bibinfo {author} {\bibfnamefont {J.}~\bibnamefont {Kono}}, \ and\ \bibinfo
  {author} {\bibfnamefont {E.}~\bibnamefont {Solano}},\ }\href@noop {}
  {\bibfield  {journal} {\bibinfo  {journal} {arXiv:1804:09275}\ } (\bibinfo
  {year} {2018})}\BibitemShut {NoStop}%
\bibitem [{\citenamefont {Scalari}\ \emph {et~al.}(2012)\citenamefont
  {Scalari}, \citenamefont {Maissen}, \citenamefont {Turcinkova}, \citenamefont
  {Hagenmuller}, \citenamefont {De~Liberato}, \citenamefont {Ciuti},
  \citenamefont {Reichl}, \citenamefont {Schuh}, \citenamefont {Wegscheider},
  \citenamefont {Beck},\ and\ \citenamefont
  {Faist}}]{scalari_ultrastrong_2012}%
  \BibitemOpen
  \bibfield  {author} {\bibinfo {author} {\bibfnamefont {G.}~\bibnamefont
  {Scalari}}, \bibinfo {author} {\bibfnamefont {C.}~\bibnamefont {Maissen}},
  \bibinfo {author} {\bibfnamefont {D.}~\bibnamefont {Turcinkova}}, \bibinfo
  {author} {\bibfnamefont {D.}~\bibnamefont {Hagenmuller}}, \bibinfo {author}
  {\bibfnamefont {S.}~\bibnamefont {De~Liberato}}, \bibinfo {author}
  {\bibfnamefont {C.}~\bibnamefont {Ciuti}}, \bibinfo {author} {\bibfnamefont
  {C.}~\bibnamefont {Reichl}}, \bibinfo {author} {\bibfnamefont
  {D.}~\bibnamefont {Schuh}}, \bibinfo {author} {\bibfnamefont
  {W.}~\bibnamefont {Wegscheider}}, \bibinfo {author} {\bibfnamefont
  {M.}~\bibnamefont {Beck}}, \ and\ \bibinfo {author} {\bibfnamefont
  {J.}~\bibnamefont {Faist}},\ }\href {\doibase 10.1126/science.1216022}
  {\bibfield  {journal} {\bibinfo  {journal} {Science}\ }\textbf {\bibinfo
  {volume} {335}},\ \bibinfo {pages} {1323} (\bibinfo {year}
  {2012})}\BibitemShut {NoStop}%
\bibitem [{\citenamefont {Todorov}\ \emph
  {et~al.}(2010{\natexlab{a}})\citenamefont {Todorov}, \citenamefont {Andrews},
  \citenamefont {Colombelli}, \citenamefont {De~Liberato}, \citenamefont
  {Ciuti}, \citenamefont {Klang}, \citenamefont {Strasser},\ and\ \citenamefont
  {Sirtori}}]{todorov_ultrastrong_2010}%
  \BibitemOpen
  \bibfield  {author} {\bibinfo {author} {\bibfnamefont {Y.}~\bibnamefont
  {Todorov}}, \bibinfo {author} {\bibfnamefont {A.~M.}\ \bibnamefont
  {Andrews}}, \bibinfo {author} {\bibfnamefont {R.}~\bibnamefont {Colombelli}},
  \bibinfo {author} {\bibfnamefont {S.}~\bibnamefont {De~Liberato}}, \bibinfo
  {author} {\bibfnamefont {C.}~\bibnamefont {Ciuti}}, \bibinfo {author}
  {\bibfnamefont {P.}~\bibnamefont {Klang}}, \bibinfo {author} {\bibfnamefont
  {G.}~\bibnamefont {Strasser}}, \ and\ \bibinfo {author} {\bibfnamefont
  {C.}~\bibnamefont {Sirtori}},\ }\href {\doibase
  10.1103/PhysRevLett.105.196402} {\bibfield  {journal} {\bibinfo  {journal}
  {Physical Review Letters}\ }\textbf {\bibinfo {volume} {105}},\ \bibinfo
  {pages} {196402} (\bibinfo {year} {2010}{\natexlab{a}})}\BibitemShut
  {NoStop}%
\bibitem [{\citenamefont {Strupiechonski}\ \emph {et~al.}(2012)\citenamefont
  {Strupiechonski}, \citenamefont {Xu}, \citenamefont {Brekenfeld},
  \citenamefont {Todorov}, \citenamefont {Isac}, \citenamefont {Andrews},
  \citenamefont {Klang}, \citenamefont {Sirtori}, \citenamefont {Strasser},
  \citenamefont {Degiron},\ and\ \citenamefont
  {Colombelli}}]{strupiechonski_sub-diffraction-limit_2012}%
  \BibitemOpen
  \bibfield  {author} {\bibinfo {author} {\bibfnamefont {E.}~\bibnamefont
  {Strupiechonski}}, \bibinfo {author} {\bibfnamefont {G.}~\bibnamefont {Xu}},
  \bibinfo {author} {\bibfnamefont {M.}~\bibnamefont {Brekenfeld}}, \bibinfo
  {author} {\bibfnamefont {Y.}~\bibnamefont {Todorov}}, \bibinfo {author}
  {\bibfnamefont {N.}~\bibnamefont {Isac}}, \bibinfo {author} {\bibfnamefont
  {A.~M.}\ \bibnamefont {Andrews}}, \bibinfo {author} {\bibfnamefont
  {P.}~\bibnamefont {Klang}}, \bibinfo {author} {\bibfnamefont
  {C.}~\bibnamefont {Sirtori}}, \bibinfo {author} {\bibfnamefont
  {G.}~\bibnamefont {Strasser}}, \bibinfo {author} {\bibfnamefont
  {A.}~\bibnamefont {Degiron}}, \ and\ \bibinfo {author} {\bibfnamefont
  {R.}~\bibnamefont {Colombelli}},\ }\href {\doibase 10.1063/1.3697660}
  {\bibfield  {journal} {\bibinfo  {journal} {Applied Physics Letters}\
  }\textbf {\bibinfo {volume} {100}},\ \bibinfo {pages} {131113} (\bibinfo
  {year} {2012})}\BibitemShut {NoStop}%
\bibitem [{\citenamefont {Dietze}\ \emph {et~al.}(2013)\citenamefont {Dietze},
  \citenamefont {Andrews}, \citenamefont {Klang}, \citenamefont {Strasser},
  \citenamefont {Unterrainer},\ and\ \citenamefont
  {Darmo}}]{dietze_ultrastrong_2013}%
  \BibitemOpen
  \bibfield  {author} {\bibinfo {author} {\bibfnamefont {D.}~\bibnamefont
  {Dietze}}, \bibinfo {author} {\bibfnamefont {A.~M.}\ \bibnamefont {Andrews}},
  \bibinfo {author} {\bibfnamefont {P.}~\bibnamefont {Klang}}, \bibinfo
  {author} {\bibfnamefont {G.}~\bibnamefont {Strasser}}, \bibinfo {author}
  {\bibfnamefont {K.}~\bibnamefont {Unterrainer}}, \ and\ \bibinfo {author}
  {\bibfnamefont {J.}~\bibnamefont {Darmo}},\ }\href {\doibase
  10.1063/1.4830092} {\bibfield  {journal} {\bibinfo  {journal} {Applied
  Physics Letters}\ }\textbf {\bibinfo {volume} {103}},\ \bibinfo {pages}
  {201106} (\bibinfo {year} {2013})}\BibitemShut {NoStop}%
\bibitem [{\citenamefont {Benz}\ \emph {et~al.}(2015)\citenamefont {Benz},
  \citenamefont {Campione}, \citenamefont {Klem}, \citenamefont {Sinclair},\
  and\ \citenamefont {Brener}}]{benz_control_2015}%
  \BibitemOpen
  \bibfield  {author} {\bibinfo {author} {\bibfnamefont {A.}~\bibnamefont
  {Benz}}, \bibinfo {author} {\bibfnamefont {S.}~\bibnamefont {Campione}},
  \bibinfo {author} {\bibfnamefont {J.~F.}\ \bibnamefont {Klem}}, \bibinfo
  {author} {\bibfnamefont {M.~B.}\ \bibnamefont {Sinclair}}, \ and\ \bibinfo
  {author} {\bibfnamefont {I.}~\bibnamefont {Brener}},\ }\href {\doibase
  10.1021/nl504815c} {\bibfield  {journal} {\bibinfo  {journal} {Nano Letters}\
  }\textbf {\bibinfo {volume} {15}},\ \bibinfo {pages} {1959} (\bibinfo {year}
  {2015})}\BibitemShut {NoStop}%
\bibitem [{\citenamefont {Askenazi}\ \emph {et~al.}(2014)\citenamefont
  {Askenazi}, \citenamefont {Vasanelli}, \citenamefont {Delteil}, \citenamefont
  {Todorov}, \citenamefont {Andreani}, \citenamefont {Beaudoin}, \citenamefont
  {Sagnes},\ and\ \citenamefont {Sirtori}}]{askenazi_ultra-strong_2014}%
  \BibitemOpen
  \bibfield  {author} {\bibinfo {author} {\bibfnamefont {B.}~\bibnamefont
  {Askenazi}}, \bibinfo {author} {\bibfnamefont {A.}~\bibnamefont {Vasanelli}},
  \bibinfo {author} {\bibfnamefont {A.}~\bibnamefont {Delteil}}, \bibinfo
  {author} {\bibfnamefont {Y.}~\bibnamefont {Todorov}}, \bibinfo {author}
  {\bibfnamefont {L.~C.}\ \bibnamefont {Andreani}}, \bibinfo {author}
  {\bibfnamefont {G.}~\bibnamefont {Beaudoin}}, \bibinfo {author}
  {\bibfnamefont {I.}~\bibnamefont {Sagnes}}, \ and\ \bibinfo {author}
  {\bibfnamefont {C.}~\bibnamefont {Sirtori}},\ }\href {\doibase
  10.1088/1367-2630/16/4/043029} {\bibfield  {journal} {\bibinfo  {journal}
  {New Journal of Physics}\ }\textbf {\bibinfo {volume} {16}},\ \bibinfo
  {pages} {043029} (\bibinfo {year} {2014})}\BibitemShut {NoStop}%
\bibitem [{\citenamefont {Askenazi}\ \emph {et~al.}(2017)\citenamefont
  {Askenazi}, \citenamefont {Vasanelli}, \citenamefont {Todorov}, \citenamefont
  {Sakat}, \citenamefont {Greffet}, \citenamefont {Beaudoin}, \citenamefont
  {Sagnes},\ and\ \citenamefont {Sirtori}}]{askenazi_midinfrared_2017}%
  \BibitemOpen
  \bibfield  {author} {\bibinfo {author} {\bibfnamefont {B.}~\bibnamefont
  {Askenazi}}, \bibinfo {author} {\bibfnamefont {A.}~\bibnamefont {Vasanelli}},
  \bibinfo {author} {\bibfnamefont {Y.}~\bibnamefont {Todorov}}, \bibinfo
  {author} {\bibfnamefont {E.}~\bibnamefont {Sakat}}, \bibinfo {author}
  {\bibfnamefont {J.-J.}\ \bibnamefont {Greffet}}, \bibinfo {author}
  {\bibfnamefont {G.}~\bibnamefont {Beaudoin}}, \bibinfo {author}
  {\bibfnamefont {I.}~\bibnamefont {Sagnes}}, \ and\ \bibinfo {author}
  {\bibfnamefont {C.}~\bibnamefont {Sirtori}},\ }\href {\doibase
  10.1021/acsphotonics.7b00838} {\bibfield  {journal} {\bibinfo  {journal} {ACS
  Photonics}\ }\textbf {\bibinfo {volume} {4}},\ \bibinfo {pages} {2550}
  (\bibinfo {year} {2017})}\BibitemShut {NoStop}%
\bibitem [{\citenamefont {Ciuti}\ \emph {et~al.}(2005)\citenamefont {Ciuti},
  \citenamefont {Bastard},\ and\ \citenamefont
  {Carusotto}}]{ciuti_quantum_2005}%
  \BibitemOpen
  \bibfield  {author} {\bibinfo {author} {\bibfnamefont {C.}~\bibnamefont
  {Ciuti}}, \bibinfo {author} {\bibfnamefont {G.}~\bibnamefont {Bastard}}, \
  and\ \bibinfo {author} {\bibfnamefont {I.}~\bibnamefont {Carusotto}},\ }\href
  {\doibase 10.1103/PhysRevB.72.115303} {\bibfield  {journal} {\bibinfo
  {journal} {Physical Review B}\ }\textbf {\bibinfo {volume} {72}},\ \bibinfo
  {pages} {115303} (\bibinfo {year} {2005})}\BibitemShut {NoStop}%
\bibitem [{\citenamefont {Todorov}\ and\ \citenamefont
  {Sirtori}(2014)}]{todorov_few-electron_2014}%
  \BibitemOpen
  \bibfield  {author} {\bibinfo {author} {\bibfnamefont {Y.}~\bibnamefont
  {Todorov}}\ and\ \bibinfo {author} {\bibfnamefont {C.}~\bibnamefont
  {Sirtori}},\ }\href {\doibase 10.1103/PhysRevX.4.041031} {\bibfield
  {journal} {\bibinfo  {journal} {Physical Review X}\ }\textbf {\bibinfo
  {volume} {4}},\ \bibinfo {pages} {041031} (\bibinfo {year}
  {2014})}\BibitemShut {NoStop}%
\bibitem [{\citenamefont {Keller}\ \emph {et~al.}(2017)\citenamefont {Keller},
  \citenamefont {Scalari}, \citenamefont {Cibella}, \citenamefont {Maissen},
  \citenamefont {Appugliese}, \citenamefont {Giovine}, \citenamefont {Leoni},
  \citenamefont {Beck},\ and\ \citenamefont
  {Faist}}]{keller_few-electron_2017}%
  \BibitemOpen
  \bibfield  {author} {\bibinfo {author} {\bibfnamefont {J.}~\bibnamefont
  {Keller}}, \bibinfo {author} {\bibfnamefont {G.}~\bibnamefont {Scalari}},
  \bibinfo {author} {\bibfnamefont {S.}~\bibnamefont {Cibella}}, \bibinfo
  {author} {\bibfnamefont {C.}~\bibnamefont {Maissen}}, \bibinfo {author}
  {\bibfnamefont {F.}~\bibnamefont {Appugliese}}, \bibinfo {author}
  {\bibfnamefont {E.}~\bibnamefont {Giovine}}, \bibinfo {author} {\bibfnamefont
  {R.}~\bibnamefont {Leoni}}, \bibinfo {author} {\bibfnamefont
  {M.}~\bibnamefont {Beck}}, \ and\ \bibinfo {author} {\bibfnamefont
  {J.}~\bibnamefont {Faist}},\ }\href {\doibase 10.1021/acs.nanolett.7b03228}
  {\bibfield  {journal} {\bibinfo  {journal} {Nano Letters}\ }\textbf {\bibinfo
  {volume} {17}},\ \bibinfo {pages} {7410} (\bibinfo {year}
  {2017})}\BibitemShut {NoStop}%
\bibitem [{\citenamefont {Paulillo}\ \emph {et~al.}(2016)\citenamefont
  {Paulillo}, \citenamefont {Manceau}, \citenamefont {Li}, \citenamefont
  {Davies}, \citenamefont {Linfield},\ and\ \citenamefont
  {Colombelli}}]{paulillo_room_2016}%
  \BibitemOpen
  \bibfield  {author} {\bibinfo {author} {\bibfnamefont {B.}~\bibnamefont
  {Paulillo}}, \bibinfo {author} {\bibfnamefont {J.-M.}\ \bibnamefont
  {Manceau}}, \bibinfo {author} {\bibfnamefont {L.~H.}\ \bibnamefont {Li}},
  \bibinfo {author} {\bibfnamefont {A.~G.}\ \bibnamefont {Davies}}, \bibinfo
  {author} {\bibfnamefont {E.~H.}\ \bibnamefont {Linfield}}, \ and\ \bibinfo
  {author} {\bibfnamefont {R.}~\bibnamefont {Colombelli}},\ }\href {\doibase
  10.1063/1.4943167} {\bibfield  {journal} {\bibinfo  {journal} {Applied
  Physics Letters}\ }\textbf {\bibinfo {volume} {108}},\ \bibinfo {pages}
  {101101} (\bibinfo {year} {2016})}\BibitemShut {NoStop}%
\bibitem [{\citenamefont {Benz}\ \emph {et~al.}(2013)\citenamefont {Benz},
  \citenamefont {Campione}, \citenamefont {Liu}, \citenamefont {Monta\~no},
  \citenamefont {Klem}, \citenamefont {Allerman}, \citenamefont {Wendt},
  \citenamefont {Sinclair}, \citenamefont {Capolino},\ and\ \citenamefont
  {Brener}}]{benz_strong_2013}%
  \BibitemOpen
  \bibfield  {author} {\bibinfo {author} {\bibfnamefont {A.}~\bibnamefont
  {Benz}}, \bibinfo {author} {\bibfnamefont {S.}~\bibnamefont {Campione}},
  \bibinfo {author} {\bibfnamefont {S.}~\bibnamefont {Liu}}, \bibinfo {author}
  {\bibfnamefont {I.}~\bibnamefont {Monta\~no}}, \bibinfo {author}
  {\bibfnamefont {J.}~\bibnamefont {Klem}}, \bibinfo {author} {\bibfnamefont
  {A.}~\bibnamefont {Allerman}}, \bibinfo {author} {\bibfnamefont
  {J.}~\bibnamefont {Wendt}}, \bibinfo {author} {\bibfnamefont
  {M.}~\bibnamefont {Sinclair}}, \bibinfo {author} {\bibfnamefont
  {F.}~\bibnamefont {Capolino}}, \ and\ \bibinfo {author} {\bibfnamefont
  {I.}~\bibnamefont {Brener}},\ }\href {\doibase 10.1038/ncomms3882} {\bibfield
   {journal} {\bibinfo  {journal} {Nature Communications}\ }\textbf {\bibinfo
  {volume} {4}} (\bibinfo {year} {2013}),\ 10.1038/ncomms3882}\BibitemShut
  {NoStop}%
\bibitem [{\citenamefont {Malerba}\ \emph {et~al.}(2016)\citenamefont
  {Malerba}, \citenamefont {Ongarello}, \citenamefont {Paulillo}, \citenamefont
  {Manceau}, \citenamefont {Beaudoin}, \citenamefont {Sagnes}, \citenamefont
  {De~Angelis},\ and\ \citenamefont {Colombelli}}]{malerba_towards_2016}%
  \BibitemOpen
  \bibfield  {author} {\bibinfo {author} {\bibfnamefont {M.}~\bibnamefont
  {Malerba}}, \bibinfo {author} {\bibfnamefont {T.}~\bibnamefont {Ongarello}},
  \bibinfo {author} {\bibfnamefont {B.}~\bibnamefont {Paulillo}}, \bibinfo
  {author} {\bibfnamefont {J.-M.}\ \bibnamefont {Manceau}}, \bibinfo {author}
  {\bibfnamefont {G.}~\bibnamefont {Beaudoin}}, \bibinfo {author}
  {\bibfnamefont {I.}~\bibnamefont {Sagnes}}, \bibinfo {author} {\bibfnamefont
  {F.}~\bibnamefont {De~Angelis}}, \ and\ \bibinfo {author} {\bibfnamefont
  {R.}~\bibnamefont {Colombelli}},\ }\href {\doibase 10.1063/1.4958330}
  {\bibfield  {journal} {\bibinfo  {journal} {Applied Physics Letters}\
  }\textbf {\bibinfo {volume} {109}},\ \bibinfo {pages} {021111} (\bibinfo
  {year} {2016})}\BibitemShut {NoStop}%
\bibitem [{\citenamefont {{Feuillet-Palma}}\ \emph {et~al.}(2012)\citenamefont
  {{Feuillet-Palma}}, \citenamefont {Todorov}, \citenamefont {Steed},
  \citenamefont {Vasanelli}, \citenamefont {Biasiol}, \citenamefont {Sorba},\
  and\ \citenamefont {Sirtori}}]{feuillet-palma_extremely_2012}%
  \BibitemOpen
  \bibfield  {author} {\bibinfo {author} {\bibfnamefont {C.}~\bibnamefont
  {{Feuillet-Palma}}}, \bibinfo {author} {\bibfnamefont {Y.}~\bibnamefont
  {Todorov}}, \bibinfo {author} {\bibfnamefont {R.}~\bibnamefont {Steed}},
  \bibinfo {author} {\bibfnamefont {A.}~\bibnamefont {Vasanelli}}, \bibinfo
  {author} {\bibfnamefont {G.}~\bibnamefont {Biasiol}}, \bibinfo {author}
  {\bibfnamefont {L.}~\bibnamefont {Sorba}}, \ and\ \bibinfo {author}
  {\bibfnamefont {C.}~\bibnamefont {Sirtori}},\ }\href {\doibase
  10.1364/OE.20.029121} {\bibfield  {journal} {\bibinfo  {journal} {Optics
  Express}\ }\textbf {\bibinfo {volume} {20}},\ \bibinfo {pages} {29121}
  (\bibinfo {year} {2012})}\BibitemShut {NoStop}%
\bibitem [{\citenamefont {Mottaghizadeh}\ \emph {et~al.}(2017)\citenamefont
  {Mottaghizadeh}, \citenamefont {Todorov}, \citenamefont {Cameau},
  \citenamefont {Gacemi}, \citenamefont {Vasanelli},\ and\ \citenamefont
  {Sirtori}}]{mottaghizadeh_nanoscale_2017}%
  \BibitemOpen
  \bibfield  {author} {\bibinfo {author} {\bibfnamefont {A.}~\bibnamefont
  {Mottaghizadeh}}, \bibinfo {author} {\bibfnamefont {Y.}~\bibnamefont
  {Todorov}}, \bibinfo {author} {\bibfnamefont {M.}~\bibnamefont {Cameau}},
  \bibinfo {author} {\bibfnamefont {D.}~\bibnamefont {Gacemi}}, \bibinfo
  {author} {\bibfnamefont {A.}~\bibnamefont {Vasanelli}}, \ and\ \bibinfo
  {author} {\bibfnamefont {C.}~\bibnamefont {Sirtori}},\ }\href {\doibase
  10.1364/OE.25.028718} {\bibfield  {journal} {\bibinfo  {journal} {Optics
  Express}\ }\textbf {\bibinfo {volume} {25}},\ \bibinfo {pages} {28718}
  (\bibinfo {year} {2017})}\BibitemShut {NoStop}%
\bibitem [{\citenamefont {Todorov}\ \emph {et~al.}(2015)\citenamefont
  {Todorov}, \citenamefont {Desfonds}, \citenamefont {Belacel}, \citenamefont
  {Becerra},\ and\ \citenamefont {Sirtori}}]{todorov_three-dimensional_2015}%
  \BibitemOpen
  \bibfield  {author} {\bibinfo {author} {\bibfnamefont {Y.}~\bibnamefont
  {Todorov}}, \bibinfo {author} {\bibfnamefont {P.}~\bibnamefont {Desfonds}},
  \bibinfo {author} {\bibfnamefont {C.}~\bibnamefont {Belacel}}, \bibinfo
  {author} {\bibfnamefont {L.}~\bibnamefont {Becerra}}, \ and\ \bibinfo
  {author} {\bibfnamefont {C.}~\bibnamefont {Sirtori}},\ }\href {\doibase
  10.1364/OE.23.016838} {\bibfield  {journal} {\bibinfo  {journal} {Optics
  Express}\ }\textbf {\bibinfo {volume} {23}},\ \bibinfo {pages} {16838}
  (\bibinfo {year} {2015})}\BibitemShut {NoStop}%
\bibitem [{\citenamefont {Todorov}\ \emph {et~al.}(2012)\citenamefont
  {Todorov}, \citenamefont {Tosetto}, \citenamefont {Delteil}, \citenamefont
  {Vasanelli}, \citenamefont {Sirtori}, \citenamefont {Andrews},\ and\
  \citenamefont {Strasser}}]{todorov_polaritonic_2012}%
  \BibitemOpen
  \bibfield  {author} {\bibinfo {author} {\bibfnamefont {Y.}~\bibnamefont
  {Todorov}}, \bibinfo {author} {\bibfnamefont {L.}~\bibnamefont {Tosetto}},
  \bibinfo {author} {\bibfnamefont {A.}~\bibnamefont {Delteil}}, \bibinfo
  {author} {\bibfnamefont {A.}~\bibnamefont {Vasanelli}}, \bibinfo {author}
  {\bibfnamefont {C.}~\bibnamefont {Sirtori}}, \bibinfo {author} {\bibfnamefont
  {A.~M.}\ \bibnamefont {Andrews}}, \ and\ \bibinfo {author} {\bibfnamefont
  {G.}~\bibnamefont {Strasser}},\ }\href {\doibase 10.1103/PhysRevB.86.125314}
  {\bibfield  {journal} {\bibinfo  {journal} {Physical Review B}\ }\textbf
  {\bibinfo {volume} {86}},\ \bibinfo {pages} {125314} (\bibinfo {year}
  {2012})}\BibitemShut {NoStop}%
\bibitem [{\citenamefont {Kittel}(1996)}]{kittel_introduction_1996}%
  \BibitemOpen
  \bibfield  {author} {\bibinfo {author} {\bibfnamefont {C.}~\bibnamefont
  {Kittel}},\ }\href@noop {} {\emph {\bibinfo {title} {Introduction to {{Solid
  State Physics}}}}},\ \bibinfo {edition} {john wiley \& sons}\ ed.\ (\bibinfo
  {year} {1996})\BibitemShut {NoStop}%
\bibitem [{\citenamefont {Cataldo}\ \emph {et~al.}(2012)\citenamefont
  {Cataldo}, \citenamefont {Beall}, \citenamefont {Cho}, \citenamefont
  {McAndrew}, \citenamefont {Niemack},\ and\ \citenamefont
  {Wollack}}]{cataldo_infrared_2012}%
  \BibitemOpen
  \bibfield  {author} {\bibinfo {author} {\bibfnamefont {G.}~\bibnamefont
  {Cataldo}}, \bibinfo {author} {\bibfnamefont {J.~A.}\ \bibnamefont {Beall}},
  \bibinfo {author} {\bibfnamefont {H.-M.}\ \bibnamefont {Cho}}, \bibinfo
  {author} {\bibfnamefont {B.}~\bibnamefont {McAndrew}}, \bibinfo {author}
  {\bibfnamefont {M.~D.}\ \bibnamefont {Niemack}}, \ and\ \bibinfo {author}
  {\bibfnamefont {E.~J.}\ \bibnamefont {Wollack}},\ }\href {\doibase
  10.1364/OL.37.004200} {\bibfield  {journal} {\bibinfo  {journal} {Optics
  Letters}\ }\textbf {\bibinfo {volume} {37}},\ \bibinfo {pages} {4200}
  (\bibinfo {year} {2012})}\BibitemShut {NoStop}%
\bibitem [{\citenamefont {Zhou}\ \emph {et~al.}(2007)\citenamefont {Zhou},
  \citenamefont {Koschny},\ and\ \citenamefont
  {Soukoulis}}]{zhou_magnetic_2007}%
  \BibitemOpen
  \bibfield  {author} {\bibinfo {author} {\bibfnamefont {J.}~\bibnamefont
  {Zhou}}, \bibinfo {author} {\bibfnamefont {T.}~\bibnamefont {Koschny}}, \
  and\ \bibinfo {author} {\bibfnamefont {C.~M.}\ \bibnamefont {Soukoulis}},\
  }\href {\doibase 10.1364/OE.15.017881} {\bibfield  {journal} {\bibinfo
  {journal} {Optics Express}\ }\textbf {\bibinfo {volume} {15}},\ \bibinfo
  {pages} {17881} (\bibinfo {year} {2007})}\BibitemShut {NoStop}%
\bibitem [{\citenamefont {Sze}\ and\ \citenamefont
  {Kwok}(2006)}]{sze_physics_2006}%
  \BibitemOpen
  \bibfield  {author} {\bibinfo {author} {\bibfnamefont {S.~M.}\ \bibnamefont
  {Sze}}\ and\ \bibinfo {author} {\bibfnamefont {K.~N.}\ \bibnamefont {Kwok}},\
  }\href@noop {} {\emph {\bibinfo {title} {Physics of {{Semiconductor
  Devices}}}}},\ \bibinfo {edition} {john wiley \& sons}\ ed.\ (\bibinfo {year}
  {2006})\BibitemShut {NoStop}%
\bibitem [{\citenamefont {Ando}\ \emph {et~al.}(1982)\citenamefont {Ando},
  \citenamefont {Fowler},\ and\ \citenamefont {Stern}}]{Ando_electronic_1982}%
  \BibitemOpen
  \bibfield  {author} {\bibinfo {author} {\bibfnamefont {T.}~\bibnamefont
  {Ando}}, \bibinfo {author} {\bibfnamefont {A.~B.}\ \bibnamefont {Fowler}}, \
  and\ \bibinfo {author} {\bibfnamefont {F.}~\bibnamefont {Stern}},\ }\href
  {\doibase 10.1103/RevModPhys.54.437} {\bibfield  {journal} {\bibinfo
  {journal} {Reviews of Modern Physics}\ }\textbf {\bibinfo {volume} {54}},\
  \bibinfo {pages} {437} (\bibinfo {year} {1982})}\BibitemShut {NoStop}%
\bibitem [{\citenamefont {Todorov}\ and\ \citenamefont
  {Sirtori}(2012)}]{todorov_intersubband_2012}%
  \BibitemOpen
  \bibfield  {author} {\bibinfo {author} {\bibfnamefont {Y.}~\bibnamefont
  {Todorov}}\ and\ \bibinfo {author} {\bibfnamefont {C.}~\bibnamefont
  {Sirtori}},\ }\href {\doibase 10.1103/PhysRevB.85.045304} {\bibfield
  {journal} {\bibinfo  {journal} {Physical Review B}\ }\textbf {\bibinfo
  {volume} {85}},\ \bibinfo {pages} {045304} (\bibinfo {year}
  {2012})}\BibitemShut {NoStop}%
\bibitem [{\citenamefont {Zanotto}\ \emph {et~al.}(2012)\citenamefont
  {Zanotto}, \citenamefont {Degl'Innocenti}, \citenamefont {Sorba},
  \citenamefont {Tredicucci},\ and\ \citenamefont
  {Biasiol}}]{zanotto_analysis_2012}%
  \BibitemOpen
  \bibfield  {author} {\bibinfo {author} {\bibfnamefont {S.}~\bibnamefont
  {Zanotto}}, \bibinfo {author} {\bibfnamefont {R.}~\bibnamefont
  {Degl'Innocenti}}, \bibinfo {author} {\bibfnamefont {L.}~\bibnamefont
  {Sorba}}, \bibinfo {author} {\bibfnamefont {A.}~\bibnamefont {Tredicucci}}, \
  and\ \bibinfo {author} {\bibfnamefont {G.}~\bibnamefont {Biasiol}},\ }\href
  {\doibase 10.1103/PhysRevB.85.035307} {\bibfield  {journal} {\bibinfo
  {journal} {Physical Review B}\ }\textbf {\bibinfo {volume} {85}},\ \bibinfo
  {pages} {035307} (\bibinfo {year} {2012})}\BibitemShut {NoStop}%
\bibitem [{\citenamefont {Liu}\ \emph {et~al.}(2016)\citenamefont {Liu},
  \citenamefont {Sun}, \citenamefont {Majumdar},\ and\ \citenamefont
  {Sorger}}]{liu_fundamental_2016}%
  \BibitemOpen
  \bibfield  {author} {\bibinfo {author} {\bibfnamefont {K.}~\bibnamefont
  {Liu}}, \bibinfo {author} {\bibfnamefont {S.}~\bibnamefont {Sun}}, \bibinfo
  {author} {\bibfnamefont {A.}~\bibnamefont {Majumdar}}, \ and\ \bibinfo
  {author} {\bibfnamefont {V.~J.}\ \bibnamefont {Sorger}},\ }\href {\doibase
  10.1038/srep37419} {\bibfield  {journal} {\bibinfo  {journal} {Scientific
  Reports}\ }\textbf {\bibinfo {volume} {6}} (\bibinfo {year} {2016}),\
  10.1038/srep37419}\BibitemShut {NoStop}%
\bibitem [{\citenamefont {Todorov}\ \emph
  {et~al.}(2010{\natexlab{b}})\citenamefont {Todorov}, \citenamefont {Tosetto},
  \citenamefont {Teissier}, \citenamefont {Andrews}, \citenamefont {Klang},
  \citenamefont {Colombelli}, \citenamefont {Sagnes}, \citenamefont
  {Strasser},\ and\ \citenamefont {Sirtori}}]{todorov_optical_2010}%
  \BibitemOpen
  \bibfield  {author} {\bibinfo {author} {\bibfnamefont {Y.}~\bibnamefont
  {Todorov}}, \bibinfo {author} {\bibfnamefont {L.}~\bibnamefont {Tosetto}},
  \bibinfo {author} {\bibfnamefont {J.}~\bibnamefont {Teissier}}, \bibinfo
  {author} {\bibfnamefont {A.~M.}\ \bibnamefont {Andrews}}, \bibinfo {author}
  {\bibfnamefont {P.}~\bibnamefont {Klang}}, \bibinfo {author} {\bibfnamefont
  {R.}~\bibnamefont {Colombelli}}, \bibinfo {author} {\bibfnamefont
  {I.}~\bibnamefont {Sagnes}}, \bibinfo {author} {\bibfnamefont
  {G.}~\bibnamefont {Strasser}}, \ and\ \bibinfo {author} {\bibfnamefont
  {C.}~\bibnamefont {Sirtori}},\ }\href {\doibase 10.1364/OE.18.013886}
  {\bibfield  {journal} {\bibinfo  {journal} {Optics Express}\ }\textbf
  {\bibinfo {volume} {18}},\ \bibinfo {pages} {13886} (\bibinfo {year}
  {2010}{\natexlab{b}})}\BibitemShut {NoStop}%
\bibitem [{\citenamefont {Amanti}\ \emph {et~al.}(2013)\citenamefont {Amanti},
  \citenamefont {Bismuto}, \citenamefont {Beck}, \citenamefont {Isa},
  \citenamefont {Kumar}, \citenamefont {Reimhult},\ and\ \citenamefont
  {Faist}}]{amanti_electrically_2013}%
  \BibitemOpen
  \bibfield  {author} {\bibinfo {author} {\bibfnamefont {M.~I.}\ \bibnamefont
  {Amanti}}, \bibinfo {author} {\bibfnamefont {A.}~\bibnamefont {Bismuto}},
  \bibinfo {author} {\bibfnamefont {M.}~\bibnamefont {Beck}}, \bibinfo {author}
  {\bibfnamefont {L.}~\bibnamefont {Isa}}, \bibinfo {author} {\bibfnamefont
  {K.}~\bibnamefont {Kumar}}, \bibinfo {author} {\bibfnamefont
  {E.}~\bibnamefont {Reimhult}}, \ and\ \bibinfo {author} {\bibfnamefont
  {J.}~\bibnamefont {Faist}},\ }\href {\doibase 10.1364/OE.21.010917}
  {\bibfield  {journal} {\bibinfo  {journal} {Optics Express}\ }\textbf
  {\bibinfo {volume} {21}},\ \bibinfo {pages} {10917} (\bibinfo {year}
  {2013})}\BibitemShut {NoStop}%
\bibitem [{\citenamefont {L\"ahnemann}\ \emph {et~al.}(2017)\citenamefont
  {L\"ahnemann}, \citenamefont {Ajay}, \citenamefont {Den~Hertog},\ and\
  \citenamefont {Monroy}}]{lahnemann_near-infrared_2017}%
  \BibitemOpen
  \bibfield  {author} {\bibinfo {author} {\bibfnamefont {J.}~\bibnamefont
  {L\"ahnemann}}, \bibinfo {author} {\bibfnamefont {A.}~\bibnamefont {Ajay}},
  \bibinfo {author} {\bibfnamefont {M.~I.}\ \bibnamefont {Den~Hertog}}, \ and\
  \bibinfo {author} {\bibfnamefont {E.}~\bibnamefont {Monroy}},\ }\href
  {\doibase 10.1021/acs.nanolett.7b03414} {\bibfield  {journal} {\bibinfo
  {journal} {Nano Letters}\ }\textbf {\bibinfo {volume} {17}},\ \bibinfo
  {pages} {6954} (\bibinfo {year} {2017})}\BibitemShut {NoStop}%
\bibitem [{\citenamefont {Feist}\ and\ \citenamefont
  {{Garcia-Vidal}}(2015)}]{feist_extraordinary_2015}%
  \BibitemOpen
  \bibfield  {author} {\bibinfo {author} {\bibfnamefont {J.}~\bibnamefont
  {Feist}}\ and\ \bibinfo {author} {\bibfnamefont {F.~J.}\ \bibnamefont
  {{Garcia-Vidal}}},\ }\href {\doibase 10.1103/PhysRevLett.114.196402}
  {\bibfield  {journal} {\bibinfo  {journal} {Physical Review Letters}\
  }\textbf {\bibinfo {volume} {114}} (\bibinfo {year} {2015}),\
  10.1103/PhysRevLett.114.196402}\BibitemShut {NoStop}%
\bibitem [{\citenamefont {Orgiu}\ \emph {et~al.}(2015)\citenamefont {Orgiu},
  \citenamefont {George}, \citenamefont {Hutchison}, \citenamefont {Devaux},
  \citenamefont {Dayen}, \citenamefont {Doudin}, \citenamefont {Stellacci},
  \citenamefont {Genet}, \citenamefont {Schachenmayer}, \citenamefont {Genes},
  \citenamefont {Pupillo}, \citenamefont {Samor\`i},\ and\ \citenamefont
  {Ebbesen}}]{orgiu_conductivity_2015}%
  \BibitemOpen
  \bibfield  {author} {\bibinfo {author} {\bibfnamefont {E.}~\bibnamefont
  {Orgiu}}, \bibinfo {author} {\bibfnamefont {J.}~\bibnamefont {George}},
  \bibinfo {author} {\bibfnamefont {J.~A.}\ \bibnamefont {Hutchison}}, \bibinfo
  {author} {\bibfnamefont {E.}~\bibnamefont {Devaux}}, \bibinfo {author}
  {\bibfnamefont {J.~F.}\ \bibnamefont {Dayen}}, \bibinfo {author}
  {\bibfnamefont {B.}~\bibnamefont {Doudin}}, \bibinfo {author} {\bibfnamefont
  {F.}~\bibnamefont {Stellacci}}, \bibinfo {author} {\bibfnamefont
  {C.}~\bibnamefont {Genet}}, \bibinfo {author} {\bibfnamefont
  {J.}~\bibnamefont {Schachenmayer}}, \bibinfo {author} {\bibfnamefont
  {C.}~\bibnamefont {Genes}}, \bibinfo {author} {\bibfnamefont
  {G.}~\bibnamefont {Pupillo}}, \bibinfo {author} {\bibfnamefont
  {P.}~\bibnamefont {Samor\`i}}, \ and\ \bibinfo {author} {\bibfnamefont
  {T.~W.}\ \bibnamefont {Ebbesen}},\ }\href {\doibase 10.1038/nmat4392}
  {\bibfield  {journal} {\bibinfo  {journal} {Nature Materials}\ }\textbf
  {\bibinfo {volume} {14}},\ \bibinfo {pages} {1123} (\bibinfo {year}
  {2015})}\BibitemShut {NoStop}%
\bibitem [{\citenamefont {Bruhat}\ \emph {et~al.}(2016)\citenamefont {Bruhat},
  \citenamefont {Viennot}, \citenamefont {Dartiailh}, \citenamefont
  {Desjardins}, \citenamefont {Kontos},\ and\ \citenamefont
  {Cottet}}]{bruhat_cavity_2016}%
  \BibitemOpen
  \bibfield  {author} {\bibinfo {author} {\bibfnamefont {L.~E.}\ \bibnamefont
  {Bruhat}}, \bibinfo {author} {\bibfnamefont {J.~J.}\ \bibnamefont {Viennot}},
  \bibinfo {author} {\bibfnamefont {M.~C.}\ \bibnamefont {Dartiailh}}, \bibinfo
  {author} {\bibfnamefont {M.~M.}\ \bibnamefont {Desjardins}}, \bibinfo
  {author} {\bibfnamefont {T.}~\bibnamefont {Kontos}}, \ and\ \bibinfo {author}
  {\bibfnamefont {A.}~\bibnamefont {Cottet}},\ }\href {\doibase
  10.1103/PhysRevX.6.021014} {\bibfield  {journal} {\bibinfo  {journal}
  {Physical Review X}\ }\textbf {\bibinfo {volume} {6}},\ \bibinfo {pages}
  {021014} (\bibinfo {year} {2016})}\BibitemShut {NoStop}%
\bibitem [{\citenamefont {Stockklauser}\ \emph {et~al.}(2017)\citenamefont
  {Stockklauser}, \citenamefont {Scarlino}, \citenamefont {Koski},
  \citenamefont {Gasparinetti}, \citenamefont {Andersen}, \citenamefont
  {Reichl}, \citenamefont {Wegscheider}, \citenamefont {Ihn}, \citenamefont
  {Ensslin},\ and\ \citenamefont {Wallraff}}]{stockklauser_strong_2017}%
  \BibitemOpen
  \bibfield  {author} {\bibinfo {author} {\bibfnamefont {A.}~\bibnamefont
  {Stockklauser}}, \bibinfo {author} {\bibfnamefont {P.}~\bibnamefont
  {Scarlino}}, \bibinfo {author} {\bibfnamefont {J.~V.}\ \bibnamefont {Koski}},
  \bibinfo {author} {\bibfnamefont {S.}~\bibnamefont {Gasparinetti}}, \bibinfo
  {author} {\bibfnamefont {C.~K.}\ \bibnamefont {Andersen}}, \bibinfo {author}
  {\bibfnamefont {C.}~\bibnamefont {Reichl}}, \bibinfo {author} {\bibfnamefont
  {W.}~\bibnamefont {Wegscheider}}, \bibinfo {author} {\bibfnamefont
  {T.}~\bibnamefont {Ihn}}, \bibinfo {author} {\bibfnamefont {K.}~\bibnamefont
  {Ensslin}}, \ and\ \bibinfo {author} {\bibfnamefont {A.}~\bibnamefont
  {Wallraff}},\ }\href {\doibase 10.1103/PhysRevX.7.011030} {\bibfield
  {journal} {\bibinfo  {journal} {Physical Review X}\ }\textbf {\bibinfo
  {volume} {7}},\ \bibinfo {pages} {011030} (\bibinfo {year}
  {2017})}\BibitemShut {NoStop}%
\bibitem [{\citenamefont {{Paravicini-Bagliani}}\ \emph
  {et~al.}(2018)\citenamefont {{Paravicini-Bagliani}}, \citenamefont
  {Appugliese}, \citenamefont {Richter}, \citenamefont {Valmorra},
  \citenamefont {Keller}, \citenamefont {Beck}, \citenamefont {Bartolo},
  \citenamefont {R\"ossler}, \citenamefont {Ihn}, \citenamefont {Ensslin},
  \citenamefont {Ciuti}, \citenamefont {Scalari},\ and\ \citenamefont
  {Faist}}]{paravicini-bagliani_magneto-transport_2018}%
  \BibitemOpen
  \bibfield  {author} {\bibinfo {author} {\bibfnamefont {G.~L.}\ \bibnamefont
  {{Paravicini-Bagliani}}}, \bibinfo {author} {\bibfnamefont {F.}~\bibnamefont
  {Appugliese}}, \bibinfo {author} {\bibfnamefont {E.}~\bibnamefont {Richter}},
  \bibinfo {author} {\bibfnamefont {F.}~\bibnamefont {Valmorra}}, \bibinfo
  {author} {\bibfnamefont {J.}~\bibnamefont {Keller}}, \bibinfo {author}
  {\bibfnamefont {M.}~\bibnamefont {Beck}}, \bibinfo {author} {\bibfnamefont
  {N.}~\bibnamefont {Bartolo}}, \bibinfo {author} {\bibfnamefont
  {C.}~\bibnamefont {R\"ossler}}, \bibinfo {author} {\bibfnamefont
  {T.}~\bibnamefont {Ihn}}, \bibinfo {author} {\bibfnamefont {K.}~\bibnamefont
  {Ensslin}}, \bibinfo {author} {\bibfnamefont {C.}~\bibnamefont {Ciuti}},
  \bibinfo {author} {\bibfnamefont {G.}~\bibnamefont {Scalari}}, \ and\
  \bibinfo {author} {\bibfnamefont {J.}~\bibnamefont {Faist}},\ }\href
  {\doibase 10.1038/s41567-018-0346-y} {\bibfield  {journal} {\bibinfo
  {journal} {Nature Physics}\ }\textbf {\bibinfo {volume} {15}},\ \bibinfo
  {pages} {186} (\bibinfo {year} {2018})}\BibitemShut {NoStop}%
\bibitem [{\citenamefont {Hagenm\"uller}\ \emph {et~al.}(2018)\citenamefont
  {Hagenm\"uller}, \citenamefont {Schachenmayer}, \citenamefont {Genet},
  \citenamefont {Ebbesen},\ and\ \citenamefont
  {Pupillo}}]{hagenmuller_intrinsic_2018}%
  \BibitemOpen
  \bibfield  {author} {\bibinfo {author} {\bibfnamefont {D.}~\bibnamefont
  {Hagenm\"uller}}, \bibinfo {author} {\bibfnamefont {J.}~\bibnamefont
  {Schachenmayer}}, \bibinfo {author} {\bibfnamefont {C.}~\bibnamefont
  {Genet}}, \bibinfo {author} {\bibfnamefont {T.}~\bibnamefont {Ebbesen}}, \
  and\ \bibinfo {author} {\bibfnamefont {G.}~\bibnamefont {Pupillo}},\
  }\href@noop {} {\bibfield  {journal} {\bibinfo  {journal} {arXiv:1810.10190}\
  } (\bibinfo {year} {2018})}\BibitemShut {NoStop}%
\bibitem [{\citenamefont {De~Liberato}\ and\ \citenamefont
  {Ciuti}(2009)}]{de_liberato_quantum_2009}%
  \BibitemOpen
  \bibfield  {author} {\bibinfo {author} {\bibfnamefont {S.}~\bibnamefont
  {De~Liberato}}\ and\ \bibinfo {author} {\bibfnamefont {C.}~\bibnamefont
  {Ciuti}},\ }\href {\doibase 10.1103/PhysRevB.79.075317} {\bibfield  {journal}
  {\bibinfo  {journal} {Physical Review B}\ }\textbf {\bibinfo {volume} {79}},\
  \bibinfo {pages} {075317} (\bibinfo {year} {2009})}\BibitemShut {NoStop}%
\end{thebibliography}%

\end{document}